\documentclass[useAMS,usenatbib]{mn2e}
\usepackage{epsfig,graphics,graphicx,bm}

\title[Secular model of migrating planet pairs]
{Modeling the secular evolution of migrating planet pairs}
\author[T.\,A. Michtchenko and A. Rodr\'iguez ]
{T.\,A. Michtchenko\thanks{E-mail: tatiana@astro.iag.usp.br} and A. Rodr\'iguez\\
Instituto de Astronomia, Geof\'{\i}sica e Ci\^encias Atmosf\'ericas, USP, Rua do Mat\~ao 1226, 05508-900 S\~ao Paulo, Brazil
}
\begin{document}

\date{Accepted ... Received ...; in original form ...}

\maketitle

\label{firstpage}

\begin{abstract}
The subject of this paper is the secular behaviour of a pair of planets evolving under dissipative forces. In particular, we investigate the case when dissipative forces affect the planetary semi-major axes and the planets move inward/outward the central star, in a process known as planet migration. To perform this investigation, we introduce fundamental concepts of conservative and dissipative dynamics of the three-body problem. Based on these concepts, we develop a qualitative model of the secular evolution of the migrating planetary pair. Our approach is based on analysis of the energy and the orbital angular momentum exchange between the two-planet system and an external medium; thus no specific kind of dissipative forces is invoked. We show that, under assumption that dissipation is weak and slow, the evolutionary routes of the migrating planets are traced by the Mode I and Mode II stationary solutions of the conservative secular problem. The ultimate convergence and the evolution of the system along one of these secular modes of motion is determined uniquely by the condition that the dissipation rate is sufficiently smaller than the proper secular frequency of the system. We show that it is possible to reassemble the starting configurations and migration history of the systems on the basis of their final states and consequently to constrain the parameters of the physical processes involved.

\end{abstract}


\section{Introduction}\label{sec0}

The secular regime of motion of multi-planetary systems is universal; in contrast with the 'accidental' resonant motion, characteristic only for specific configurations of the planets, secular motion is present everywhere in phase space, even inside the resonant region. The secular behaviour of a pair of planets evolving under dissipative forces is the principal subject of this paper, particularly, the case when the dissipative forces affect the planetary semi-major axes and the planets move inward/outward the central star, the process known as planet migration. It is presently accepted that migration plays an important role in the dynamical history of planetary systems. Several mechanisms have been proposed to be responsible for planet migration; some of these are i) planet interactions with a protoplanetary disk of gas/dust, ii) gravitational scattering and clearing of remnant planetesimal debris by the  planets, iii) direct collisions between the planets, and iv) tidal interactions of the planets with a central star. There exists a vast literature on this issue; the reader is referred to Armitage (2010) for reviews of planet migration and references therein.

The study of planet migration is frequently connected to theories of planet formation. The modeling of interactions of growing planets with protoplanetary discs is a high-complexity task independently whether it is done analytically or in the form of numerical simulations. Despite the significative progress stimulated by the discovery of extra-solar planets, the studies still suffer from various limitations (see Lubow and Ida (2010) for the recent review). The main obstacle is probably a lack of knowledge about the structure of planet-forming gaseous discs and their detailed physical properties. The huge list of unknown parameters requests numerous numerical simulations, characterized by high computational costs, and introduces uncertainties in the results obtained. Together with poorly determined physical properties of the process, unknown starting configurations of the planets complicate investigations in the cases of migration of the tidally affected close-in planets and scattering the planetesimals in the late stage of the planetary formation.

In this paper, we purpose a simple method for a qualitative study of the planet migration, which does not require a  detailed knowledge of dissipative mechanism, its physical properties and starting configuration of the migrating system.  This is because we do not consider any specific kind of dissipative forces (they can be originated by tidal torques in the disc, tidal interactions with the star, ejection of planetesimals by planets, direct collisions etc).  Knowing that, under dissipation, the energy and the orbital angular momentum of the planet system are no longer conserved, we model the migration through the variation of the energy and parameters of the secular dynamics: the angular momentum and the semi-major axes ratio. Similar approaches have already been used successfully in dynamical studies (see Malhotra 1995, Murray et al. 2002, Nelson and Papaloizou 2002, among others).

Exploring basic concepts of the conservative and dissipative dynamics, we develop a model of secular dynamics of a migrating pair of planet in Section \ref{sec3}. We first analyze possible effects of dissipation on the global dynamical quantities of the secular system, such as energy and orbital angular momentum, which are invariable during the conservative motion. From these effects, the most known is the increase/decrease of the orbital energy of the system defined by the Keplerian motions of the planets. The result is the variation of the semi-major axes and the consequent expansion/contraction of the planetary orbits.

The effects of dissipative forces on the secular component of the total energy of the system have been studied much less. The secular energy is defined by mutual secular interactions of the planets and its dissipation affects mainly the planet eccentricities provoking their damping oscillations. Even when only one planet is directly subjected to friction forces, the orbital angular momentum exchange between the planets affects the eccentricity of the other planet. We show that, if the dissipation is sufficiently weak and slow (adiabatical approach), the planets ultimately evolve into a nearly steady state defined by a stationary solution of the conservative problem, either Mode I or Mode II of motion (Michtchenko and Ferraz-Mello 2001), depending on the physical parameters of the system. In what follows, the already damped system continues to evolve under acting dissipative forces, tracing the family formed by the stationary solutions. We show that the final configuration of the system is defined by the balance between the orbital and secular energy variations during migration, together with the angular momentum exchange between the system and an external medium.

The developing of the model of the migrating pair of the planets follows essentially two steps: The first one is the calculation of the families composed of the Mode I and Mode II stationary solutions, for all possible values of planetary semi-major axes and angular momentum of the system. Parameterized by the mass ratio, these families will define the evolutionary routes of the migrating system in phase space.  To obtain the families of stationary solutions, with no restrictions on the values of planet eccentricities, we use the precise semi-analytical approach developed in (Michtchenko and Malhotra 2004).

The second step consists of the investigation of the stability of Mode I and Mode II of motion, to determine which of these modes will play a role of the center, toward which the system is attracted during migration. We show that the stability of a center depends on the combined effect of three conditions: The first one is related to the law of the orbital angular momentum exchange between the system and the external medium. For instance, it can imply that the orbital decay of the planets results in the decreasing of the eccentricities of their orbits and vice versa. The second condition is related to the loss/gain of the secular energy by the system. It should be noting that the gain of the secular energy results in divergent planetary orbits, while the loss of the secular energy in convergent orbits. The third condition relates two parameters of the secular problem, the mass ratio $m_2/m_1$ and the semi-major axes ratio $a_1/a_2$ (indices $1$ and $2$ correspond to the inner and outer planets, respectively). We show that Mode I of motion plays a role of an attractive center, when planetary orbits diverge and $\sqrt{a_1/a_2} < m_2/m_1$, or, when planetary orbits converge and $\sqrt{a_1/a_2} > m_2/m_1$. In the opposite cases, the Mode II of motion is stable. We show that an immediate consequence of this result is that the evolutionary history of a dissipative system can not be accessed through the back in time simulations of its evolution.

In order to illustrate the performance of our model, we do a series of purely numerical experiments, whose results are presented in Section \ref{sec4}. We apply the model to several fictitious systems, varying the planetary masses and semi-major axes and simulating both types of orbits, divergent and convergent ones. We initially assume that only the inner planet is affected by external dissipative forces and that the orbital angular momentum of the system is conserved during migration. It should be noted that these assumptions describe well the secular interaction of a tidally affected synchronously rotating planet with its outer companion.

We test then the dependence of the migration evolution on the starting conditions of the secular system. We verify that the final configuration of the system is independent on where in the phase space the system started its secular evolution, provided that the starting values of eccentricities satisfy the conservation of the orbital angular momentum. Thus, despite a qualitative nature of our model, it is able to indicate precisely the domains in phase space where the system could start its dissipative evolution. We also investigate the dependence of the simulated planetary paths on the rate of the migration process and the individual masses of the planets and determine the applicability limits of the  adiabatical approach.

In the next section, we extend the model to the case when the system exchanges orbital angular momentum with the external medium. We introduce the ${\rm AM}$-leakage, defined as a portion ($\alpha$) of the angular momentum variation produced by migration (i.e. the expansion/cotraction of the planetary orbits), which is extracted (or added to) from the system. We construct the migration routes of the system evolving with different values of the factor $\alpha$. This approach provide us with a general idea of how the system could evolve under a variety of migration conditions and physical models.

In Section \ref{sec8}, the model is adapted to the case when only the motion of the outer planet is affected by dissipative forces. Analyzing the migration tracks obtained in this case, together with the previously obtained ones, we show that the evolution and the final state of the secular system, evolving under weak dissipative forces, with typical values of $\alpha$ smaller than 1, depend on the balance between the orbital and secular energy variations, as follows:

- the loss of the orbital and secular energy defines the total circularization of two convergent orbits, allowing a capture in low-order mean-motion resonances;

- the loss of the orbital energy and the gain of the secular energy define the totally circularized orbits and the increasing mutual planetary distance;

- the gain of the orbital and secular energy defines the increasing mutual planetary distance, together with the increase of the eccentricities of the orbits;

- the gain of the orbital energy and the loss of the secular energy define the continuous increase of the eccentricities of two convergent orbits, when the secular system is broken due some external perturbations, such as mean-motion resonance interactions or close encounters between the planets.

In Discussions, we show several examples found among the planets of the Solar System and the known extra-solar systems, whose actual configurations could contain records of the dissipative evolution in the past, according to one of the above scenarios. As a final consideration, we discuss advantages and disadvantages of the model developed in this paper.

Fundamentals needed to understand the developed model are given in Appendices \ref{sec1} and \ref{sec2}. In Appendix \ref{sec1} we discuss the main features of the conservative secular motion, developing the first-order planar Hamiltonian model of the three-body problem. We show that: i) it is linear when the motion is projected on a sphere (Pauwels 1983); ii) the planet mass  and semi-major axes ratios, together with the angular momentum of the system, play the role of free parameters of the secular problem; iii) the eccentricities of the planetary orbits oscillate in anti-phase around one of two stable stationary solutions, known as Mode I and Mode II of motion. Recall that Mode I characterizes the aligned configuration of the planetary orbits, when the averaged mutual distance between the planets is minimal and consequently the energy of the system is also minimal. On contrary, Mode II characterizes the anti-aligned configuration and the maximal energy of the system, when the averaged mutual distance between the planets is maximal.

In Appendix \ref{sec2}, we study the behaviour of a linear secular system moving under action of dissipative forces.
We show that the behaviour of the system is similar to the behaviour of the well-studied harmonic oscillator moving under friction (see Andronov et al. 1966, Chap. I). The anti-phase oscillations of the planet eccentricities around a fixed center are damped and the system ultimately evolves into a steady state defined by a stable focus. We show that the stable focus of the secular system is close to the center defined either Mode I or Mode II stationary solutions of the conservative problem, depending on the physical parameters of the system and dissipation. We note that this result is independent on the specific form of the dissipative force, but requires only sufficiently small and slow dissipation of the energy.

\section{The model}\label{sec3}

\subsection{The conservative secular evolution}
First we consider a conservative three-body system consisting of the central star ($M$) and two planets with masses $m_1$ and $m_2$ (hereafter, the indexes $i=1$, $2$ correspond to the inner and outer planets, respectively). In the heliocentric reference frame, the canonical set of variables is constituted by the relative position vectors $\vec{r}_i$ and conjugate momenta $\vec{p}_i=m_i\,\frac{{\rm d}\vec{\rho}_i}{{\rm d}t}$, where the position vectors $\vec{\rho}_i$ are relative to the center of gravity of the three-body system. If only mutual gravitational interactions are taking into account, the Hamiltonian of the problem is written in the form
\begin{equation}\label{eq:1-1}
{\mathcal H}=\underbrace{\sum_{i=1}^2 (\frac{\vec{p}^{\,2}_i}{m^\prime_i}
-\frac{\mu_i\,m_i^\prime}{|\vec{r}_i|})}_{{\rm Keplerian\,\,\,\, part}}
-\underbrace{k^2\frac{m_1\,m_2}{\Delta}}_{{\rm direct\,\,\,\, part}}
+\underbrace{\frac{(\vec{p}_1\times\vec{p}_2)}{M}}_{{\rm indirect\,\,\,\, part}},
\end{equation}
where $k^2$ is Gaussian constant, $\mu_i=k^2(M+m_i)$, $m_i^\prime=m_i\,M/(M+m_i)$ and $\Delta = |\vec{r}_1-\vec{r}_2|$. The first term is the sum of Keplerian motions of the planets and the second and third terms produce direct and indirect perturbations among the planets, respectively.

We suppose that the planets move in the same plane; then the Hamiltonian (\ref{eq:1-1}) can be written, in terms of canonical elliptic variables and the disturbing function $R$, as
\begin{equation}\label{eq:1-2}
{\mathcal H}=-\sum_{i=1}^2 \frac{\mu_i^2\,m_i^{\prime\,3}}{2\,L_i^2} -\frac{k^2\,m_1\,m_2}{a_2}\times
R(L_i,I_i,\lambda_i,\varpi_i),
\end{equation}
where the mass-weighted Poincar\'e variables are introduced as
\begin{equation}\label{eq:1-3}
 \begin{array}{rrccc}
\lambda_i & {\rm mean\,\,longitude,}             & L_i & = &m_i^\prime\sqrt{\mu_ia_i}\\
-\varpi_i & {\rm longitude\,\,of\,\,perihelion,} & I_i & = &L_i(1-\sqrt{1-e_i^2}).
\end{array}
\end{equation}
$a_i$ and $e_i$ stand, respectively, for the semi-major axes and the eccentricities of the planetary orbits. In order to investigate the secular behaviour of the system, we apply the averaging procedure to the Hamiltonian (\ref{eq:1-2}). The averaging done with respect to the mean longitudes of the planets, removes from the problem the short periodic oscillations. The averaged Hamiltonian is defined by
\begin{equation}
\begin{array}{lll}
{\overline{\mathcal H}}&=& {\mathcal H}_{\rm{Kep}} + {\mathcal H}_{\rm{sec}}\\
             &=& -\sum_{i=1}^2 \frac{\mu_i^2\,m_i^{\prime\,3}}{2\,L_i^2}\\
             & &-  \frac{1}{{(2\pi)^2}}\int_0^{2\pi}\int_0^{2\pi}\frac{k^2\,m_1\,m_2}{a_2} \, R(L_i,I_i,\lambda_i,\varpi_i)\,d\lambda_1d\lambda_2,
\end{array}
\label{eq:Hamilnumeric}
\end{equation}
where the averaging procedure need be done over only the direct part of the disturbing function, because the indirect part does not contain secular terms (Brouwer and Clemence, 1961).

After the elimination of the short periodic terms, the averaged Hamiltonian does not depend on $\lambda_i$; consequently, the semimajor axes of the planet orbits, $a_1$ and $a_2$, are constant and serve simply as parameters in the Hamiltonian (\ref{eq:Hamilnumeric}). In addition, due to the D'Alembert's rule, the $\varpi$-dependence in the averaged disturbing function is only through $\Delta\varpi=\varpi_2-\varpi_1$; this implies that $I_1+I_2$ is also a constant of motion. This quantity is so-called angular momentum deficit of the system and, up to second order in masses, can be written as
\begin{eqnarray}
{\rm AMD}& = &\sum_{i=1}^2 m_i\,n_i\,a_i^2 \left( 1 - \sqrt{1-e_i^2} \right ){\rm .}
\label{eq:AMD}                                                                                                            \end{eqnarray}
In contrast with the orbital angular momentum of the system given by
\begin{equation}\label{eq:AM}
{\rm AM}  = \sum_{i=1}^2 m_i\,n_i\,a_i^2\,\sqrt{1-e_i^2},
\end{equation}
${\rm AMD}$ is invariable solely in the secular problem, due to the fact that, in this problem, $a_1$ and $a_2$ are parameters.

It is worth emphasizing that, since the secular interactions do not depend on the positions of the planets on the orbits, the secular behaviour of the system can be described by gravitational interactions between two coplanar rings, each one with the mass of the corresponding planet (Wu and Goldreich 2002). The massive rings do not change their sizes under secular perturbations ($a_i=$ const), but only their shapes, defined by $e_i$, and the mutual orientation, defined by the secular angle $\Delta\varpi$.  The detailed description of the conservative secular dynamics of the two-planet system is given in Appendix \ref{sec1}.

The energy of the system given by the Hamiltonian (\ref{eq:Hamilnumeric}) is conserved during the secular evolution. For the purpose of this work, it can be re-written in analogy with the expression (\ref{eq:1-1}) as:
\begin{equation}\label{eq:energy}
 \overline{{\mathcal H}} = -\frac{\mu_1m^\prime_1}{2a_1} -\frac{\mu_2m^\prime_2}{2a_2}- k^2\frac{m_1\,m_2}{D},
\end{equation}
where the indirect part of the disturbing function was omitted. In the above equation, the two Keplerian terms are dominant; they are functions solely of the parameters $a_i$ and, consequently, are constant in time. Thus, due to the conservation of the energy of the system, the latter term must be also conserved during the evolution.  The constant $D$ is introduced as
\begin{equation}\label{eq:D}
D=-k^2\,m_1\,m_2/{\mathcal H}_{\rm sec},
\end{equation}
where ${\mathcal H}_{\rm sec}$ is  the secular part of the Hamiltonian given in Equation (\ref{eq:Hamilnumeric}), in general form, and in Equation (\ref{eq:hamil}), in the linear approximation.  Despite the fact that $D$ is a complicated function of the planet semi-major axes and eccentricities, it can be used as a convenient measure of the separation between two orbits. For instance, in the case of two circular orbits, $D=a_2/b^{(0)}_{1/2}$, where $b^{(0)}_{1/2}$ is the Laplace  coefficient which is an increasing function of the increasing semi-major axes ratio $a_1/a_2$ (Callegari et al. 2004).

\subsection{The dissipative secular  evolution}
Now we suppose that two planets, interacting secularly (the planets are far enough from any mean-motion resonance), undergo external perturbations which affect the energy and angular momentum of the system. One consequence of such perturbations is that the planet semi-major axes are altered and the planets move inward/outward the central star, the process known as planet migration. The variation of the energy (\ref{eq:energy}) over a small time increment is given by
\begin{equation}\label{eq:delta-E}
\Delta \overline{{\mathcal H}} = \frac{\mu_1m^\prime_1\Delta a_1}{2a^2_1} + \frac{\mu_2m^\prime_2\Delta a_2}{2a^2_2}+ \frac{k_2m_1m_2\Delta D}{D^2},
\end{equation}
where $\Delta a_1$, $\Delta a_2$ and $\Delta D$ are variations of the semi-major axes and the separation between the orbits, respectively.  The sum of the two first terms defines the change of {\it the orbital (Keplerian) energy} of the system, while the latter term defines the change of {\it the secular energy} of the system. In what follows, we will show that the balance between both defines the migration path of the system in phase space.  It is worth noting  that the variation of the secular energy can occur even when the semi-major axes are not affected by external forces, but when eccentricities are changed (see Appendix \ref{sec2-1}).

For the sake of simplicity, we suppose initially that only the semi-major axis of the inner planet $a_1$ is affected. The orbital decay of the close-in inner planet due to tidal interactions with the central star can be approximated by this assumption. (The case of the change of $a_2$ will be discussed in Section \ref{sec8}.)

According to Equation (\ref{eq:delta-E}), the loss of orbital energy of the inner planet ($\Delta a_1 < 0$) does not affect the orbital energy of its companion; thus, the semi-major axis of the outer planet remains constant during migration. In the other words,  there is no exchange of the orbital energy between the planets in the secular approximation (Wu and Goldreich 2002). The coupling between two orbits is defined by the secular term in Equation (\ref{eq:delta-E}); it is responsible for transfer of the tidally induced changes in the orbital elements of the inner planet to the outer planet (for details see Appendix \ref{sec2-2}). It is worth noting that, during divergent migration, when $\Delta a_1<0$ (or $\Delta a_2>0$), the system gains the secular energy, since the separation between two orbits is increased ($\Delta D > 0$). Inversely, during convergent migration, when $\Delta a_1 > 0$ (or $\Delta a_2 < 0$), $\Delta D < 0$ and the system loses the secular energy.

We further assume that the orbital angular momentum of the system is conserved during migration. In the approximation that only the inner planet is affected by tidal interactions with the star, this assumption is valid when the planet rotation has been synchronized and effects of the stellar tides are not considered (the case of tidal interactions of a close-in planet with a very slow rotating central star) (Casenave et al. 1980, Correia and Laskar 2010, Rodr\'iguez et al. 2011a). In this case, the decrease $\Delta a_1$ produces the damping of the inner planet eccentricity as follows:
\begin{equation}
\Delta e^{\rm{ex}}_1 = \frac{(1-e_1^2)}{2a_1e_1}\Delta a_1,
\label{condition}\end{equation}
where the index "${\rm ex}$" indicates that this eccentricity variation is a response to external forces and it must be added to the variation originated by the secular interaction with the other planet. Note that according to the above equation, under the assumption of invariable $\textrm{AM}$, the orbital decay is halted when the planet orbit is circularized. It should be emphasized that the hypothesis of the invariable orbital angular momentum is introduced here to facilitate our understanding of the dissipative behaviour of the system; in Section \ref{sec7}, the model will be extended to more general case, when the angular momentum of the system is exchanged with the external medium.

In contrast with the orbital energy, there exists the exchange of the orbital angular momentum between the planets. From Equation (\ref{eq:AM}), we can express, up to second order in masses, the coupling variations of both eccentricities as
\begin{equation}\label{eq:e2}
\Delta e_2= - \frac{m_1}{m_2}\sqrt{\frac{a_1}{a_2}}\frac{e_1}{e_2}\sqrt{\frac{1-e_1^2}{1-e_2^2}}\Delta e_1.
\end{equation}
The above equation shows that, in addition to the secular anti-phase variations of both eccentricities, the orbital decay of the inner planet ($\Delta a_1$) and the eccentricity damping $\Delta e^{\rm{ex}}_1$ (\ref{condition}) provoke the smooth decrease of $e_2$ (see for details Rodr\'iguez et al. 2011b).

The secular behaviour of the eccentricities under acting friction forces is analyzed in detail in Appendix \ref{sec2}. We show that during the orbital decay of the inner planet,  both eccentricities present damped oscillations; when the amplitude of the oscillation of $e_1$ tends to zero, the amplitude of $e_2$ also tends to zero. The only condition needed is that the dissipation rate is sufficiently slow, compared to the proper period of the secular oscillation. The final configuration of the damped system is either aligned or anti-aligned orbits, depending on the sign of $\Delta e^{\rm{ex}}_1$ and the physical parameters of the system, the mass and semi-major axes ratios. Seen in phase space (e.g., Figure \ref{figure4}), the damped system moves along the spiral trajectory attracted to the stable center, which can be either Mode I (aligned orbits) or Mode II (anti-aligned orbits) stationary solution of the conservative problem, characterized by the maximal and minimal energy, respectively.  This state of the system is often referred to as 'quasi-stationary' in the literature (Mardling 2007, Batygin et al. 2009).
\begin{figure}
\def\capfrac{1}
\centerline{\hbox{
\includegraphics[clip,width=0.45\textwidth]{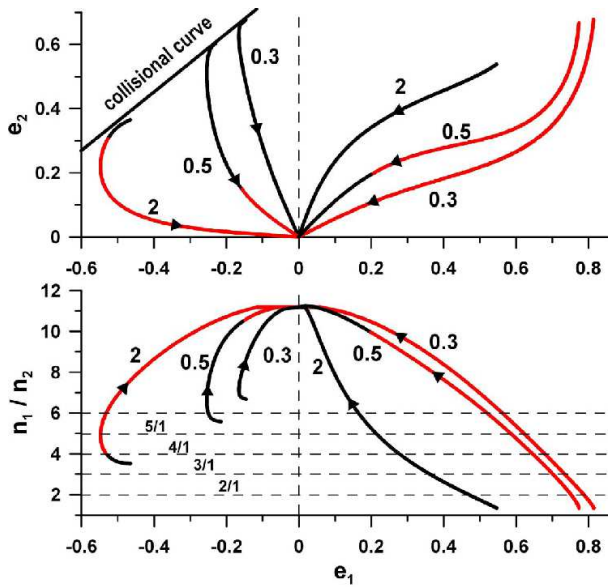}
}}
\caption{The families of stationary solutions (Mode I and Mode II), parameterized by the mass ratio $m_2/m_1$ (values beside each curve). 
Top: the ($e_1$,$e_2$) representative plane. Bottom: the ($e_1$,$n_1/n_2$)--plane. On both panels, $\Delta\varpi$ is fixed at $0$ along positive values on the $e_1$-axis) or $180^\circ$ along negative values on the $e_1$-axis). The branches of the curves in black color corresponds to solutions with $I^Z_1 < I^Z_2$, and in red color to solutions with $I^Z_1 > I^Z_2$, where $I^Z_1$ and $I^Z_2$ are the partial angular momentum deficits of the planets. Arrows show the direction of divergent orbits.
}
\label{figure6}
\end{figure}

During the evolution, the attracting center itself dislocates slowly in the phase space, due to changes of the secular parameter $a_1/a_2$ and the consequent change of the orbits separation $D$ (see an example in Figure \ref{figure4-3}). Even when 'quasi-stationary' state is attained, the system continues to evolve under action of dissipative forces, following closely the Mode I or Mode II family of stationary solutions of the conservative problem (some of these solutions are shown in Figure \ref{figure6}). The smooth migration of the system is halted only when either both orbits are circularized, in the case of $\Delta a_1 < 0$, or the system enters into the domain of phase space where secular model is not longer valid, in the case of $\Delta a_1 > 0$. In the former case, $\Delta e^{\rm{ex}}_1 = \Delta a_1 =0$, when $e_1=0$, according to Equation (\ref{condition}). In the latter case, the planets approach each other sufficiently to introduce strong short-terms or resonance perturbations which are not considered by the secular theory.

\subsection{Evolutionary routes}
As described above, the families of the conservative stationary solutions draw migration routes of the secular systems evolving under friction forces. Three of those families are shown in Figure \ref{figure6}. The families were calculated using the algorithm described in Michtchenko and Malhotra (2004), for continuous values of the ratio $a_1/a_2$ and for three different values of the mass ratio $m_2/m_1$, namely $2.0$, $0.5$ and $0.3$.  Along each family ${\rm AM}$ is fixed at the value calculated for the system given by $a_1=0.02$\,AU, $a_2=0.1$\,AU and $e_1=e_2=0.01$.

We plot the families on two different planes:  the ($e_1$,$e_2$)--plane of the planet eccentricities (top panel) and the ($e_1$,$n_1/n_2$)--plane (bottom panel), where the ratio of the mean motions of the planets $n_1/n_2$ is computed from the ratio of their semi-major axes through the third law of Kepler. On both panels, $\Delta\varpi$ is fixed: at $0$, along positive values on the $e_1$-axis, or at $180^\circ$, along negative values on the $e_1$-axis.

Each family consists of two distinct branches connected at the origin. The branch with positive $e_1$-values corresponds to Mode I of motion, while the branch with negative $e_1$-values corresponds to Mode II. Which from these modes will play a role of the stable center and, consequently, trace the trajectory of the migrating system  is discussed in detail in Section \ref{sec2-3}. As shown, the stability of the center is defined by the relation between $I_1^Z$ and $I_2^Z$, which are the partial angular momentum deficits of the planets, evaluated at the center. For $\Delta a_1<0$ (and $\Delta e^{\rm{ex}}_1 < 0$), the system evolves into the center with $I_1^Z < I_2^Z$. This is also true for $\Delta a_2>0$ (and $\Delta e^{\rm{ex}}_2 > 0$). In both cases (orbital decay of the inner planet and orbital expansion of the outer planet) the planets move away from each other. The orbits are diverging, their separation $D$ is increasing,  and the system gains the secular energy, according to Equation (\ref{eq:delta-E}).
\begin{figure}
\def\capfrac{1}
\centerline{\hbox{
\includegraphics[clip,width=0.5\textwidth]{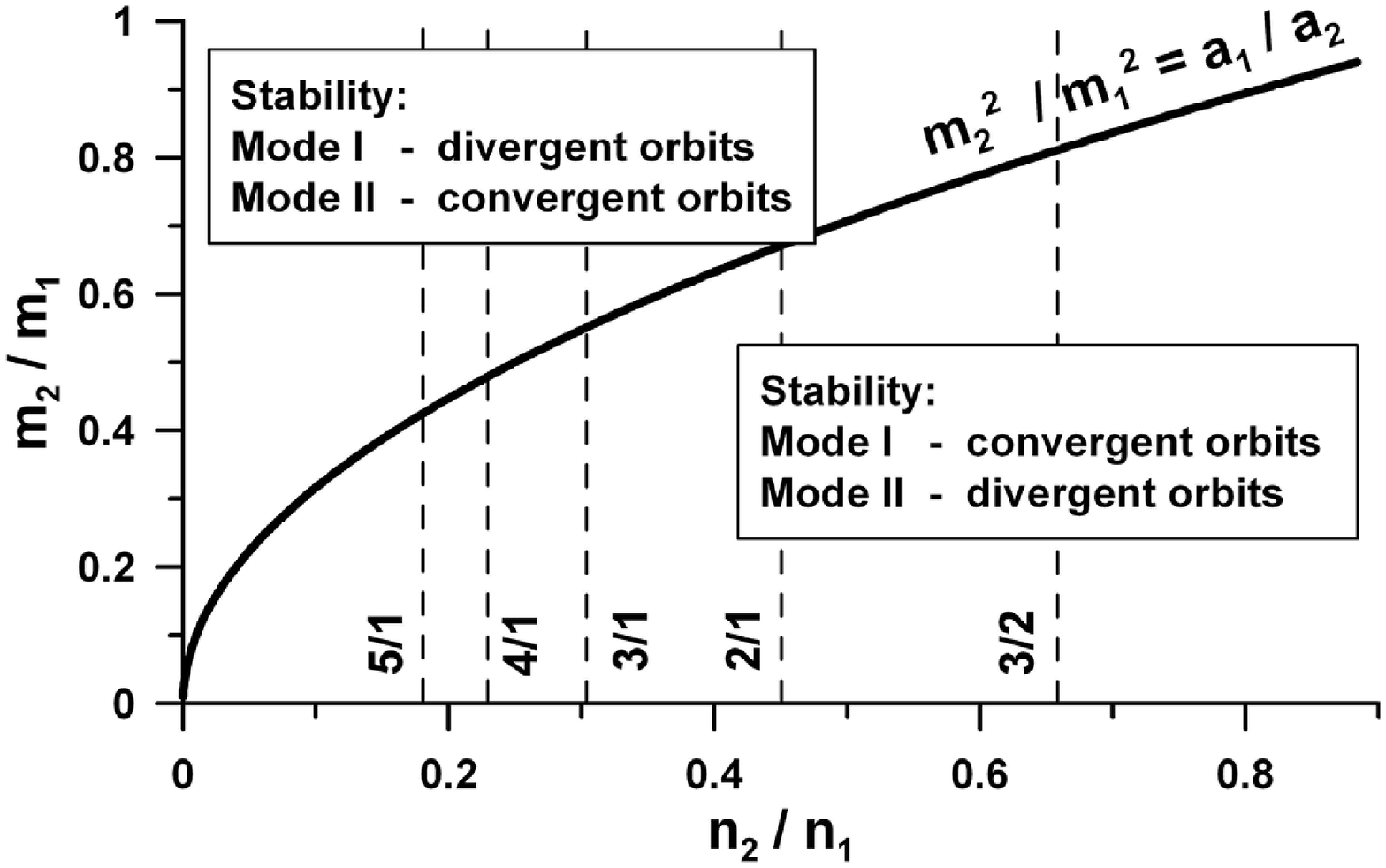}
}}
\caption{Domains of the stable centers defined by the condition (\ref{condition1}) on the parametric ($n_2/n_1$,$m_2/m_1$)--plane, where $n$ is the mean motion of the planet.  This case corresponds to the condition (\ref{eq:sgn}); in the opposite case, the stability of the domains is inverted. Locations of the main mean-motion resonances are indicated.}
\label{figure7}
\end{figure}

The situation is inverse for convergent orbits, when $\Delta a_1>0$ or $\Delta a_2<0$. In this case, the separation $D$ is decreasing and the system loses the secular energy. As a consequence, it evolves into the center with $I_1^Z > I_2^Z$. The transition condition $I_1^Z = I_2^Z$ can be re-written, up to second order in masses and eccentricities,  as
\begin{equation}
\sqrt{\frac{a_1}{a_2}} = \frac{m_2}{m_1}.
\label{condition1} \end{equation}

In Figure \ref{figure6}, the centers with $I_1^Z < I_2^Z$ are plotted by black dots and, those with $I_1^Z > I_2^Z$, by red dots. According to described above, for divergent orbits, the Mode I of motion is stable for $m_2/m_1=2.0$, while the Mode II is stable for $m_2/m_1=0.3$. The transition value of the mass ratio given by the condition (\ref{condition1}) is equal to $0.447$, for $a_1/a_2=0.2$. The family of stationary solutions parameterized by $m_2/m_1=0.5$ in Figure \ref{figure6} is close to the transition case. We can observe that, in this case, the centers change the stability during the evolution of the system along the branch. Thus, in order to access the stability of the center, when $m_2/m_1$ is close to the critical value, the condition (\ref{condition2}), which takes into account planetary eccentricities, must be applied.

It is worth emphasizing that the determination of stability of the center was done assuming that $\Delta a_1$ and $\Delta e_1$ have the same sign (see Equation (\ref{condition})), that is a consequence of the conservation of the orbital angular momentum of the system during migration. In the more general case, we introduce the condition
\begin{equation}\label{eq:sgn}
\rm{sgn}(\Delta a_{1,2})=\rm{sgn}(\Delta e^{\rm{ex}}_{1,2})
\end{equation}
and resume the information on the stability of the centers in Figure \ref{figure7}, where we plot the curve given by Equation (\ref{condition1}), on the parametric plane ($n_2/n_1$, $m_2/m_1$). If the condition (\ref{eq:sgn}) is satisfied, the divergent orbits ultimately converge to the Mode I stationary solutions in the domain above the curve. In the same domain, the convergent orbits tend to the Mode II solutions.  In the domain below the characteristic curve, the stability of the secular modes is opposite.  If the condition (\ref{eq:sgn}) is not satisfied, the stability shown in Figure \ref{figure7} is just inverted. It should be noted that the condition (\ref{eq:sgn}) is typical for the known dissipative processes in the planetary systems. The dynamical interpretation of this condition, from the point of view of the orbital angular momentum exchange, will be discussed in Section \ref{sec7}.

According to our model, the system moving under slow dissipation evolves following the stationary solutions of the conservative secular system, which draw the evolutionary route that the planets acquired from their initial configurations towards the present locations. Three such families parameterized by different mass ratios were shown in Figure \ref{figure6}. Analyzing these routes and migration conditions, we can i) do robust predictions about where in the phase space the system could arrive from, and ii) conjecture on the possible final configurations  of the system. For instance, in the case of inner planet migration, we conclude that, when the system loses its orbital energy and gains the secular one, its evolution is characterized by the divergent migration and decreasing eccentricities, with final configurations on the totally circularized orbits. The value of $a_1$ has a lower limit which can be obtained from Equation (\ref{eq:AM}), for given $\textrm{AM}$ and $a_2$ and fixing $e_1=e_2=0$; it is given in Equation (\ref{eq:a1-min}).

During the convergent migration, when system gains the orbital energy and loses the secular one, the evolution is characterized by increasing eccentricities; the final state of the system is uncertain, since it approaches the strong mean-motion resonances and close encounters domain, where short-period perturbations could disrupt the system. It should be emphasized that the described scenarios are possible when the migration condition (\ref{eq:sgn}) is satisfied. In the opposite  case, the behaviour of the eccentricities is inverted. Finally, the similar analysis with respect of the migrating outer planet will be done in Section \ref{sec8}.
\begin{figure}
\def\capfrac{1}
\centerline{\hbox{
\includegraphics[clip,width=0.45\textwidth]{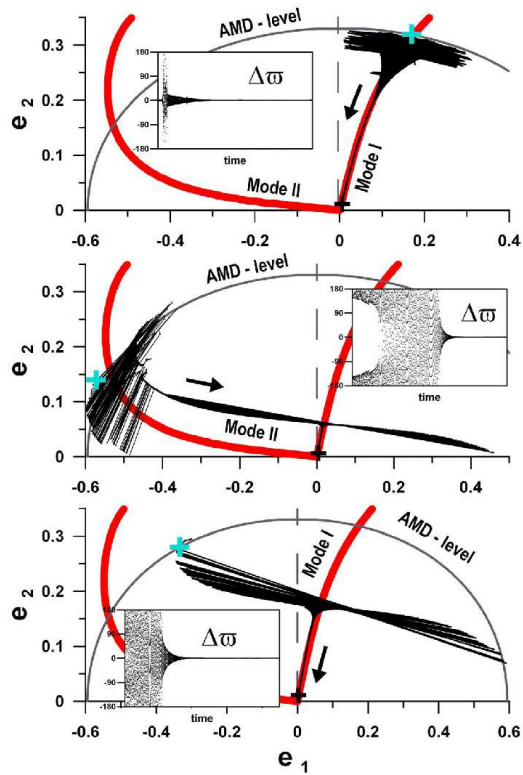}
}}
\caption{Dependence of the migrating evolution on initial conditions on the ($e_1$, $e_2$)--plane. Three different starting positions of the system are shown by cyan symbols. The final (current) configuration of the system is shown by a black symbol. Families of stationary solutions are plotted by thick red curves. Mode I corresponds to positive $e_1$--values, while Mode II to negative $e_1$--values. Arrows show the direction of the evolution of the system. The mass ratio is fixed at $m_2/m_1=2.0$.}
\label{figure8-1}
\end{figure}

\section{Numerical simulations}\label{sec4}

To illustrate the behaviour of the two-planet secular system evolving under dissipative forces, we perform a series of numerical simulations. For this task, we use a purely numerical approach which consists of integrations of exact equations of motion averaged through the applied on-line low-pass filter (Michtchenko et al. 2002). We do not invoke any specific kind of dissipative forces; the orbital migration is modeled in the exact equations of motion of the planets through constant perturbations in their semi-major axes and eccentricities, in analogy with Malhotra (1994) and Nelson and Papaloizou (2002).

The system under study consists of the central star with mass $M=1.0\,M_{\rm Sun}$ and two gravitationally interacting planets.  Different values of the planetary masses were used in this work; they will be specified for each experiment reported in this section. We place the system in its current configuration described by the semi-major axes $a_1=0.02$\,AU and $a_2=0.1$\,AU and small eccentricities $e_1=e_2=0.01$. In this section we assume that the dissipative forces are acting only on the inner planet. We also assume that the orbital angular momentum of the system is conserved (the more general case of the $\textrm{AM}$ transfer will be studied in Section \ref{sec7}) and calculate its value for the current configuration of the system. Thus, during the simulation, the inner semi-major axis is forced to decay by $\Delta a_1$, after each time increment (typically smaller than one tenth of the orbital period), and the value of $e_1$ is corrected by the amount given in (\ref{condition}). For the sake of simplicity, in this work we adopt a linear regime of the orbital decay, in the form $<\dot{a}_1> =$ const, where const may be positive or negative.

It should be emphasizing that the model allows us to study separately the secular regime of motion of the system. In fact, the averaging procedure shown in Equation (\ref{eq:Hamilnumeric}) removes  all effects of short-period terms and of mean-motion commensurabilities. This is not a case of the numerical integrations of the exact equations of motion, whose output contains information on all possible regimes of motion, which are overlapping intrinsically. This fact should be kept in mind when the results obtained by both approaches are compared.

\subsection{Dependence on initial conditions}

We first test the dependence of the migration evolution of the secular system on its initial configuration. We fix the set of the planetary masses at $m_1=1.0\,M_{\rm J}$ and $m_2=2.0\,M_{\rm J}$, and choose  arbitrarily the starting semi-major axes as $a_1=0.0311$\,AU and $a_2=0.1$\,AU. These values correspond to $n_1/n_2=5.76$ and place the system far from any strong mean-motion resonance.

The description of the starting configuration of the system must be completed by the initial values of $e_1$, $e_2$ and $\Delta\varpi$. Their choice is illustrated in Figure \ref{figure8-1}, where we show three representative planes ($e_1$,$e_2$), each one presenting the different starting position of the system. Since the orbital angular momentum is conserved in this experiment, all possible initial values of eccentricities are located on the ${\rm AMD}$--level obtained, for given planetary masses and initial $a_1$ and $a_2$, from Equation (\ref{eq:AMD}); this level is shown on all graphs in Figure \ref{figure8-1}, together with the families of stationary solutions plotted by thick red curves.

The first set of the eccentricities (cyan symbol on the top graph) was chosen in the neighborhood of the Mode I solution, with the initial $\Delta\varpi=0$. The second set (cyan symbol on the middle graph) was chosen in the opposite configuration, in the vicinity of the Mode II solution, with the initial $\Delta\varpi=180^\circ$. The third set of the initial eccentricities (cyan symbol on the bottom graph) was in the intermediate  configuration, with the initial value of $\Delta\varpi$ within the interval from 0 to $360^\circ$.

The chosen initial conditions were numerically propagated in time, according to the described above procedure, with the constant  dissipation rate $<\dot{a}_1> = -10^{-9}$AU/yr; the resulting trajectories are plotted in the ($e_1$,$e_2$)--planes  by solid curves in Figure \ref{figure8-1}. In the box, on each panel, we show the time evolution  during migration of the secular angle $\Delta\varpi$.

Figure \ref{figure8-1} shows that, independently, where the system starts the evolution, the divergent orbits, with $\Delta a_1<0$, ultimately evolve into the Mode I of motion ($\Delta\varpi=0$), which guides the system towards its current position in the vicinity of the origin. This result, obtained for the system whose parameters satisfy the condition $\sqrt{a_1/a_2} < m_2/m_1$, is in complete accordance with our model described in Section \ref{sec3}.

The important conclusion from the performed experiment is that the final state of the migrating secular system is independent of its initial position in the phase space, defined by given values of the parameters: the mass and semi-major axes ratios and the orbital angular momentum. As a consequence, to assess the system configuration in the past, we have to know solely the starting value of $a_1$, while the values of both eccentricities and of the secular angle may be chosen arbitrarily, with only restriction that the eccentricities belong to the $\textrm{AMD}$-level defined by the masses and the semi-major axes of the planets. Inversely, the obtained in such a way magnitudes of the eccentricities could impose constrains on the possible starting values of $a_1$.
\begin{figure*}
\def\capfrac{1}
\centerline{\hbox{
\includegraphics[clip,width=1.0\textwidth]{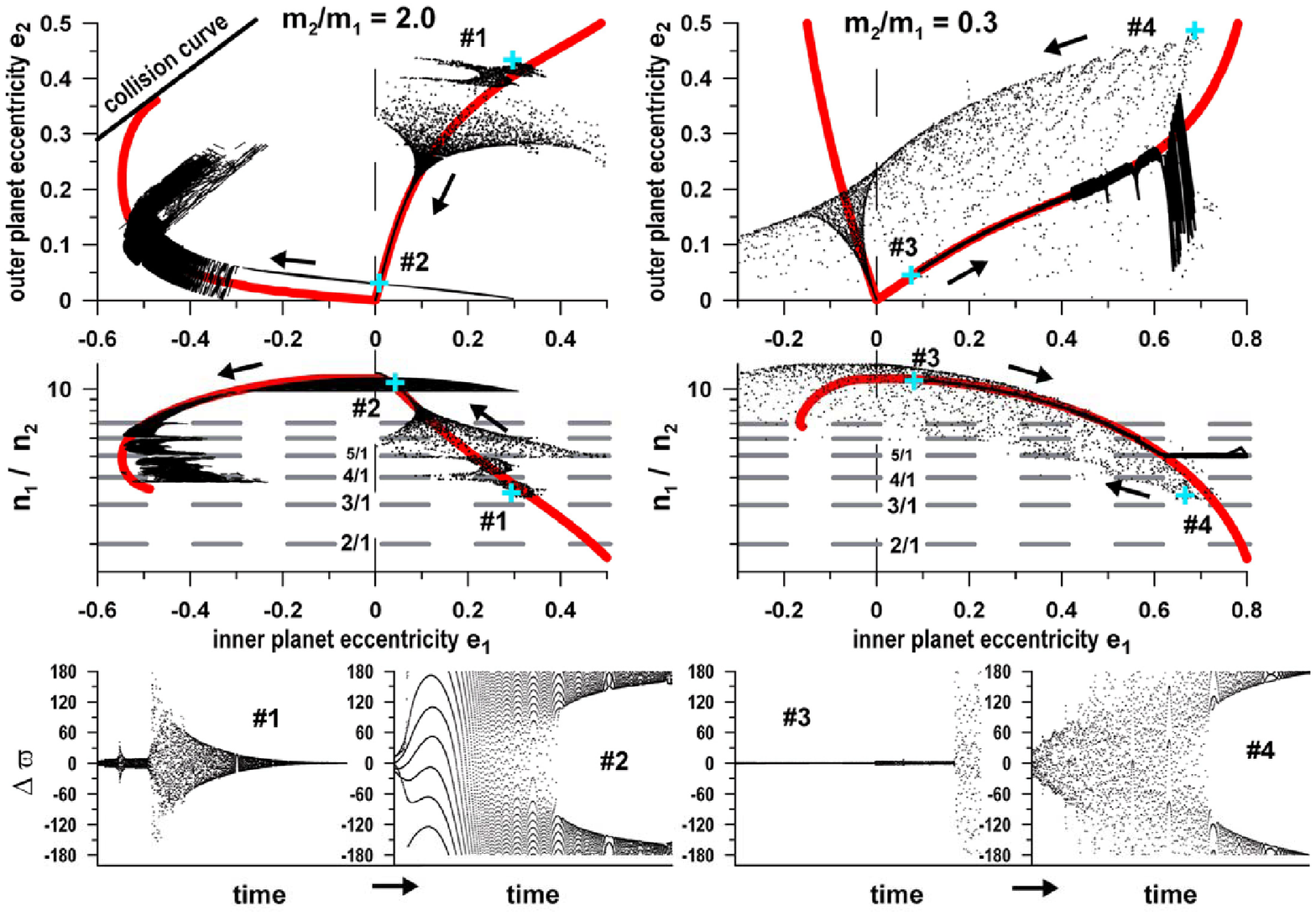}
}}
\caption{Numerical simulations of the migrating two-planet system. Families of stationary solutions are plotted by thick red curves. Mode I corresponds to positive $e_1$--values, while Mode II to negative $e_1$--values. Starting positions of the system are shown by cyan symbols and indicated by the corresponding run number. Runs \#1 and \#4 are shown by dots, while \#2 and \#3 are shown by solid lines. Arrows beside each path show the direction of the evolution of the system. }
\label{figure8}
\end{figure*}

\subsection{Dependence on the mass and semi-major axes ratios}

In this section we test the stability of the domains shown in Figure \ref{figure7}. For this task, we vary the masses and semi-major axes of the planets and consider both types of migration, divergent and convergent. The output of the four performed runs is shown in Figure \ref{figure8} on the representative planes ($e_1$,$e_2$) (top row) and ($e_1$,$n_1/n_2$) (middle row). On each plane we plot the family of stationary solutions obtained for the planet mass ratio used in the simulation; the mass ratio values are shown on the top of each column. The time evolutions of the secular angle $\Delta\varpi$ are shown on the bottom panels, each one indicating the number of the corresponding run.

\subsubsection{The case of $\sqrt{a_1/a_2} < m_2/m_1$}

The first run (\#1) simulates the migration of the system composed of two planets with the masses $m_1=1.0\,M_{\rm J}$ and $m_2=2.0\,M_{\rm J}$. The starting inner semi-major axis was $a_1=0.0455$\,AU, which corresponds to $n_1/n_2=3.24$.  Since the final configuration of the system is independent on the initial eccentricities, their values were chosen arbitrarily on the initial $\textrm{AMD}$--level, in the vicinity of the Mode I solution,  with initial $\Delta\varpi=0$. The dissipation was simulated with the constant rate $<\dot{a}_1>=-10^{-9}$AU/yr, which induces the divergent orbits and the gain of the secular energy.

The resulting path is shown by black dots on the graphs from the left column in Figure \ref{figure8}. Its starting position shown by a cyan symbol is indicated by \#1. For given values of the masses and semi-major axes, the condition $\sqrt{a_1/a_2} < m_2/m_1$ is always satisfied during the evolution of the system. In this case, according to our model, the dissipation of the energy causes the damped eccentricity oscillations and the convergence of the system to the Mode I of motion. Consequently, this secular mode drives the system toward the origin, where both orbits are circularized. This scenario is fully reproduced by the simulation \#1. The time evolution of the secular angle $\Delta\varpi$ shown on the bottom panel in Figure \ref{figure8}, clearly illustrates the capture in the Mode I of motion.

The observable deviations of the planetary path from the Mode I of motion is due to passages through several mean-motion resonances during migration. During these passages the planet eccentricities are strongly excited, but they are damped again as soon as the system leaves the mean-motion resonance. The correlation between the passages through the mean-motion resonances and excitations of the eccentricities can be clearly observed on the ($e_1$,$n_1/n_2$)--plane in Figure \ref{figure8}\,{\it middle}. Far from the main mean-motion resonances, only the component of the smooth decrease can be observed in the evolution of the planetary eccentricities.

It is worth noting that, in the theory of oscillations, the capture into the quasi-steady state is often referred to as 'relaxation', to indicate the return of the system to equilibrium after being perturbed by an external force. In this work, we do not use this term, because there is no qualitative difference between finite-amplitude oscillations around the stable center and nearly zero-amplitude oscillations (and smooth evolution) in the very close vicinity of the center.

Now we suppose that the position marked by \#2 on the top and middle panels in Figure \ref{figure8}\,{\it left column}, is the current position of the system. This configuration can be used as the starting point to simulate the migration of the same pair of planets, evolving now under the positive variation of the inner semi-major axis, with the rate $<\dot{a}_1>=10^{-9}$AU/yr. When the orbit of the inner planet is expanded, the evolution of the system is convergent and the secular energy is lost. Note that the run \#2 may be considered as a back in time simulation of the run \#1.

The trajectory obtained is shown by a continuous line on the graphs in Figure \ref{figure8}\,{\it left column}. Starting on the nearly circular orbits with aligned periastra, the planet eccentricities are exited very quickly. The increasing amplitudes of oscillations lead to the circulation of the secular angle $\Delta\varpi$, whose behaviour is shown on the bottom panel in Figure \ref{figure8}. After a stage of circulatory evolution, the periastra of the orbits enter in the anti-aligned regime of motion, when the system converges to the Mode II of motion. The system continues to evolve along this secular mode in the direction of increasing eccentricities, crossing several mean-motion resonances, until the convergent orbits reach the strong 4/1 resonance.  The chaotic effects of this resonance on the high-eccentricity planet motion cause the disruption of the system.

We verify that the result obtained is in perfect accordance with our model, which predicts that the convergent orbits, satisfying the condition $\sqrt{a_1/a_2} < m_2/m_1$, will ultimately converge to and follow the Mode II of motion (see Figure \ref{figure7}). It is worth noting that the evolution of the trajectory \#2 is completely distinct from that of the run \#1, despite that the run \#2 is just a back in time simulation of the run \#1.  Thus, the important conclusion from the done experiment is that, for dissipative systems, forward and backward in time integrations are qualitatively different; in the other words, the evolutionary history of dissipative systems can not to be studied through integrations going back in time.
\begin{figure}
\def\capfrac{1}
\centerline{\hbox{
\includegraphics[clip,width=0.5\textwidth]{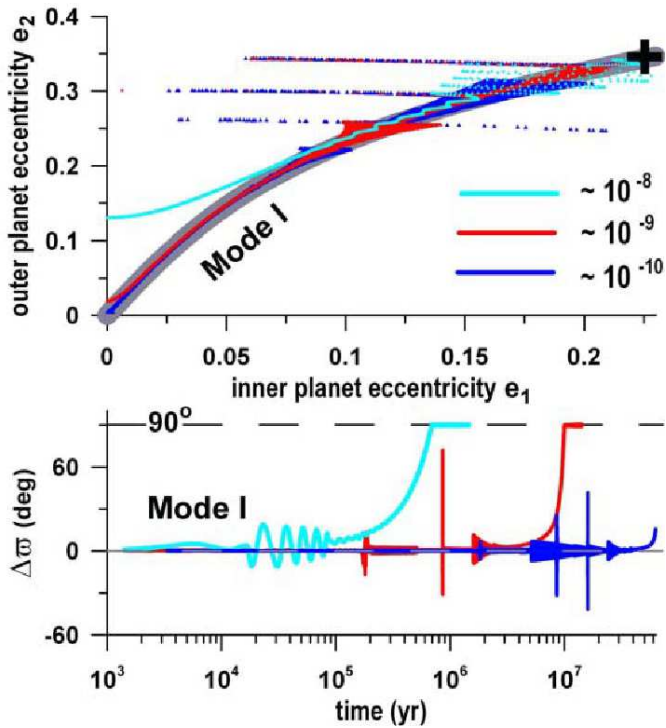}
}}
\caption{Dissipative evolution of the system along Mode I family of stationary solutions, for three different values of the dissipation rate. The mass ratio $m_2/m_1$ is fixed at 3 and all orbits start at the same initial configuration, shown by a black symbol. The excitation due to the passages through main mean-motion resonances and damped oscillations around Mode I can be clearly observed.}
\label{figure11}
\end{figure}

\subsubsection{The case of $\sqrt{a_1/a_2} > m_2/m_1$}

According to our model illustrated in Figure {\ref{figure7}}, the stability of the centers shown in the previous section is inverted in this case. By definition, $a_1 < a_2$ always; thus the above condition is satisfied only for the systems with a less massive outer companion. For this reason, two simulations, \#3 and \#4, shown on the graphs from the right column in Figure \ref{figure8}, set the planetary masses at $m_1=1.0\,M_{\rm J}$ and $m_2=0.3\,M_{\rm J}$. The position of the Mode I and Mode II stationary solutions was recalculated for a given mass ratio and is shown by the thick red curves on the corresponding graphs.

The initial conditions of the run \#3 were chosen similar to those of the run \#2: the system started in the vicinity of the Mode I, with the positive dissipation rate, $<\dot{a}_1>=10^{-9}$AU/yr, resulting in convergent orbits and in the loss of the secular energy by the system. The planetary path obtained is shown by solid lines in the top and middle planes in Figure \ref{figure8}{\it right column}.

In analogy with the evolution of the path \#2, the system moves in the direction of high eccentricities and of small $n_1/n_2$. The smooth evolution of the system is interrupted by the passages through the main mean-motion resonances, when the planet eccentricities are strongly excited. When the high-eccentricity system approaches the low-order 5/1 resonance, it is broken out. The only contrast with the path \#2 is that, at condition $\sqrt{a_1/a_2} > m_2/m_1$, the motion follows closely the Mode I family, that can be clearly seen on the bottom graph, where time variation of the secular angle $\Delta\varpi$ is shown.

The run \#4 was calculated with the dissipation rate $<\dot{a}_1>=-10^{-9}$AU/yr and with initial conditions similar to those of the run \#1. It is shown by black dots in Figure \ref{figure8}{\it right column}.  In contrast with the path \#1, in this case, the divergent planetary eccentricities suffer strong excitations. The behaviour of the secular angle $\Delta\varpi$ on the bottom panel in Figure \ref{figure8} shows that the system leaves quickly the oscillation motion around Mode I and evolves into the circulatory regime of motion. The resulting path of the run \#4 shows the capture of the system into the steady state of the Mode II and the consequent circularization of the planetary orbits. Once again, the comparison of the results of the runs \#3 and \#4 show the inadequacy of the simulation of the past of dissipative dynamical systems through backward in time integrations.
\begin{figure}
\def\capfrac{1}
\centerline{\hbox{
\epsfig{figure=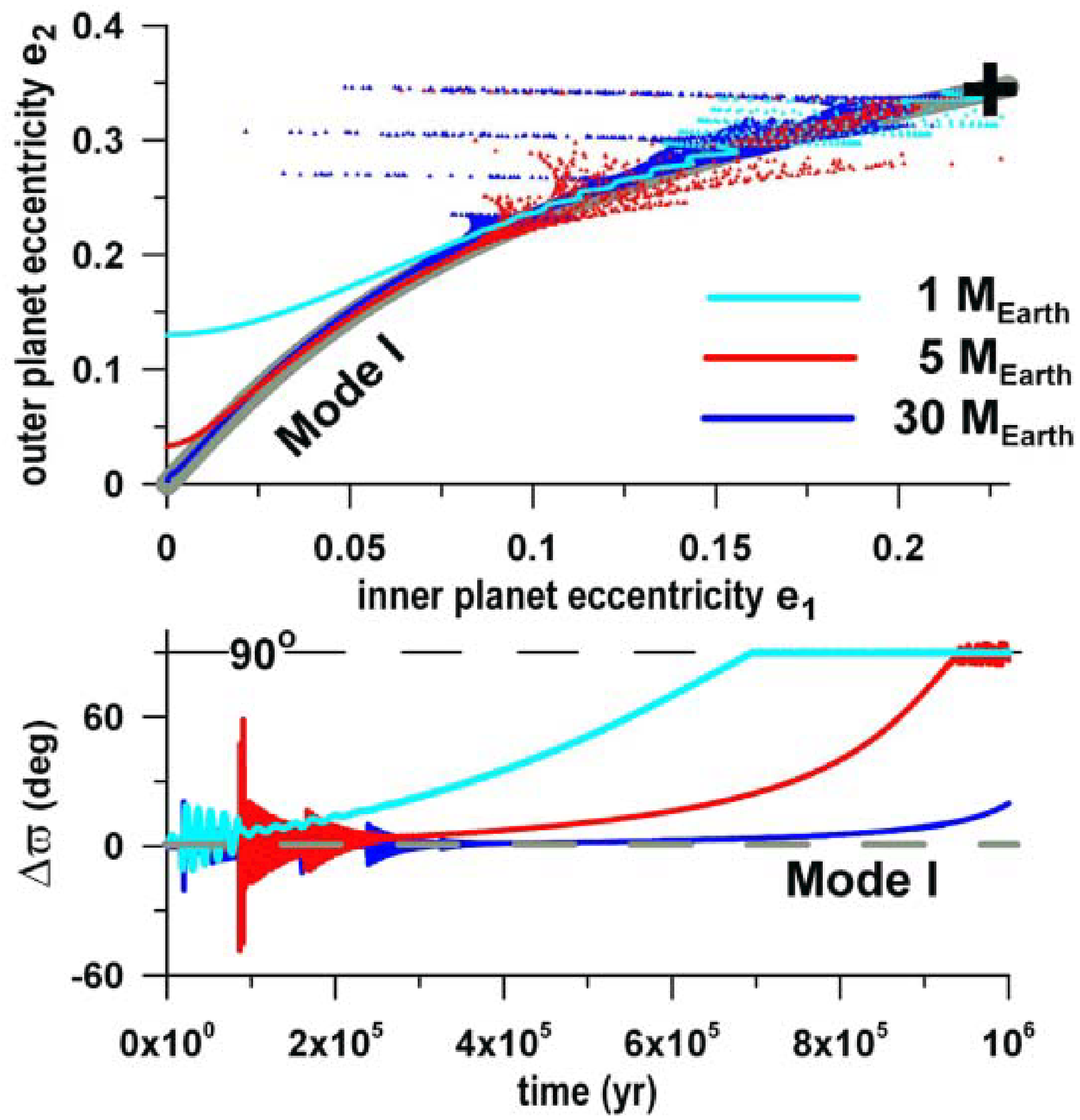, height=9cm,angle=0}
}}
\caption{Dissipative evolution of the system along Mode I family of stationary solutions, for three different values of the individual mass of the inner planet. The mass ratio is fixed at 3, the dissipation rate is fixed at $-3\times10^{-8}$\,AU/yr and all orbits start at the same initial configuration, shown by a black symbol. The excitation due to the passages through main mean-motion resonances is more strong for more massive planets. The deviation from the conservative case is more important for less massive planets.}
\label{figure12}
\end{figure}

\subsection{Dependence on the dissipation rate}\label{sec5}

In this experiment, we simulate the migrating motion of the planets using different values of the dissipation rate. The physical parameters of the system and its initial configuration are same in all simulations: the masses are fixed at $m_1=1.0\,M_{\rm{Earth}}$ and $m_2=3.0\,M_{\rm{Earth}}$, the semi-major axes at $a_1=0.0455$\,AU and $a_2=0.1$\,AU, and the system starts in the vicinity of the Mode I of motion, with $\Delta\varpi=0$. The only parameter which is different in each simulation is the dissipation rate of the orbital decay of the inner planet, $<\dot{a}_1>$, which was kept constant during each simulation.

Figure \ref{figure11} shows the planetary paths on the representative plane ($e_1$,$e_2$) (top panel) and the time evolution of the secular angle $\Delta\varpi$ (bottom panel). The different colors correspond to the different values of the dissipation rate: cyan color to $-3\times10^{-8}$\,AU/yr, red color to $-3\times10^{-9}$\,AU/yr and blue color to $-3\times10^{-10}$\,AU/yr. The thick gray curve represents the stationary solutions of the Mode I of motion. The starting configuration of all orbits is shown by a black symbol.

According to our model, the migrating system will converge to and follow the Mode I family of stationary solutions, under the condition that the dissipation is sufficiently slow. This behaviour is observed in Figure \ref{figure11}, which shows also deviations of the planet paths from the predicted route (thick gray curve). The vertical spreading of the points is originated by the excitation of eccentricities during the passages through mean-motion resonances and have been discussed in the previous section.

Our attention is focused on the monotonous deviations of the migrating paths from the Mode I family, observed at low eccentricities. This type of deviation is clearly related to the magnitude of the  dissipation rate: faster is dissipation, larger is the deviation. The final configuration of the rapidly migrating system consists of the circular orbit of the inner planet, {\it the non-circular} orbit of the outer planet (top panel in Figure \ref{figure11}) and the secular angle $\Delta\varpi$, which is seen to be 'captured' at $90^\circ$ (bottom panel). This behaviour of the system is associated to the phenomenon of {\it the aperiodic damping}, which occurs when the dissipation rate exceeds the proper frequency of the system (see Section \ref{sec2-1}). It is worth observing that, if the migration is stopped after the system has attained its final configuration,  the planets start the circulatory motion around a corresponding center belonging to the family of the stationary solutions.

\begin{figure}
\def\capfrac{1}
\centerline{\hbox{
\epsfig{figure=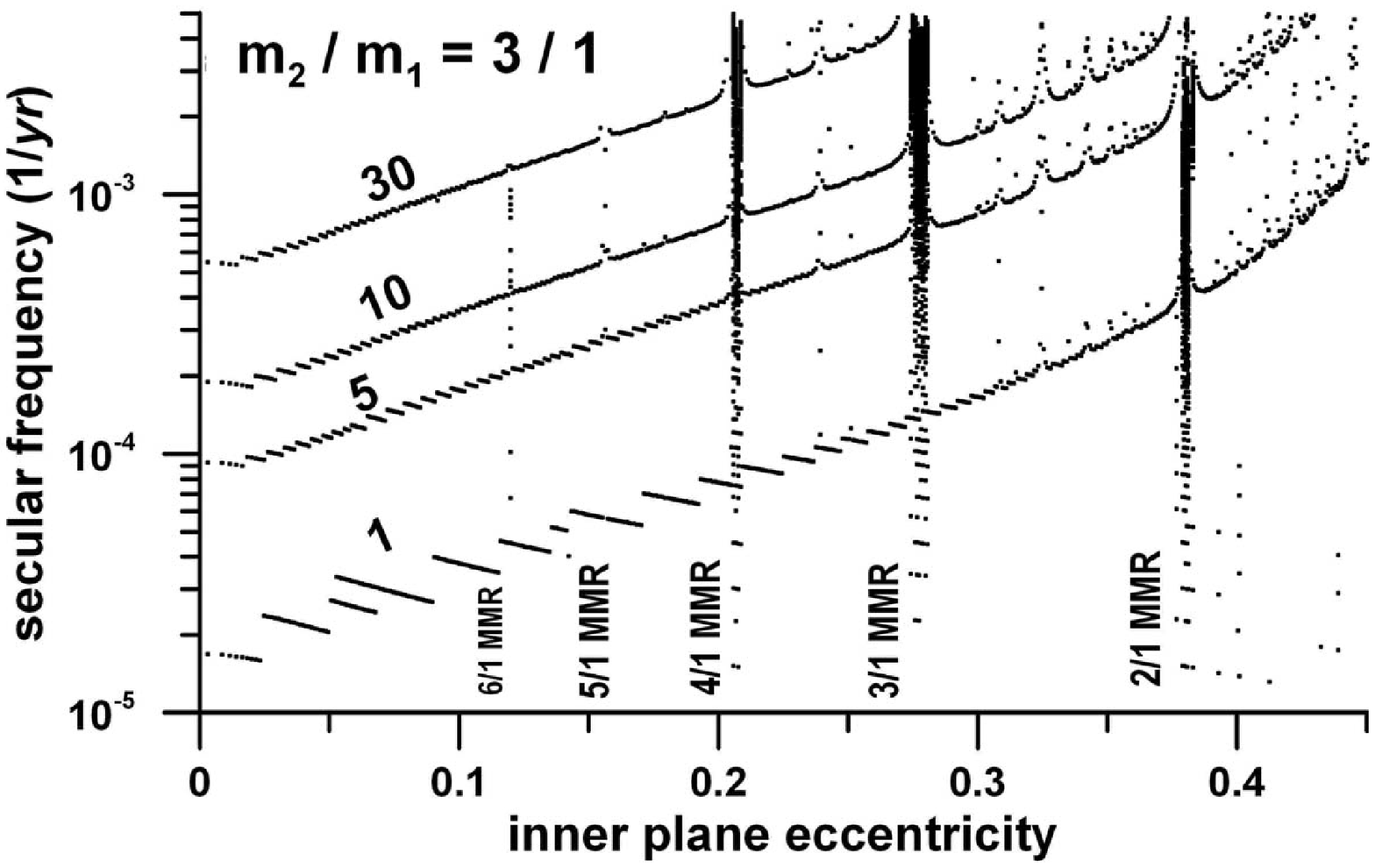, height=5.5cm,angle=0}
}}
\caption{The proper secular frequency calculated along the Mode I family shown in Figure \ref{figure12}, in the logarithmic scale. The mass ratio $m_2/m_1$ is kept fixed at 3, but the individual planetary masses, initially fixed at $m_1=1\,M_{\rm Earth}$ and $m_2=3\,M_{\rm Earth}$, were multiplied by factors 5, 10 and 30. The values of the mass factor are indicated beside each curve. The vertical bands are due to perturbations during the passages through the main mean-motion resonances (MMR).}
\label{figure9}
\end{figure}

As will be shown in the next section, the proper frequency of the secular system decreases following a power law, as the eccentricities decrease (Figure \ref{figure9}). During the rapid decay of the inner planet, the magnitude of the proper frequency may become comparable to that of the dissipation rate. In this case, the system deviates strongly from the conservative stationary solutions and can even evolve into the singularity at $\Delta\varpi=90^\circ$, as shown in Figure \ref{figure5} in Section \ref{sec2-1}. We can clearly observe these effects in the evolution of the secular angle  $\Delta\varpi$ on the bottom panel in Figure \ref{figure11}. It should be emphasizing that, in this work, we present only the qualitative analysis of the effects of the aperiodic damping on the orbits of the migrating planets. The detailed study of this process requires an introduction of a specific migration mechanism and is out of the scope of this paper.

\subsection{Dependence on the individual planetary masses}\label{sec6}

In this section we perform several simulations fixing the mass ratio, but varying individual planetary masses. The mass ratio is fixed at $m_2/m_1=3.0$, while the dissipation rate $<\dot{a}_1> $ is fixed at $-3\times10^{-8}$\,AU/yr. All simulations start at the same initial configuration, close to the Mode I stationary solution, at $a_1=0.0455$\,AU and $a_2=0.1$\,AU. According to our model, for given mass and semi-major axes ratios, this migrating system will evolve along the secular Mode I of motion (see Figure \ref{figure7}).

Figure \ref{figure12} shows the planetary paths obtained: with the individual masses fixed at $m_1=1\,M_{\rm{Earth}}$ and $m_2=3\,M_{\rm{Earth}}$ by cyan color, with these masses multiplied by factor 5 by red color, and multiplied by factor 30 by blue color. The correlation between the planet masses and the monotonous deviation from the Mode I family (thick gray curve) is clear: smaller is the mass factor, larger is the deviation of the migrating path from the route traced by the Mode I stationary solutions. This dependence can be easily understood from Figure \ref{figure9}, where we plot the proper frequencies obtained along the Mode I family. (The proper frequency is defined by the angular velocity of the $\Delta\varpi$--precession.) The frequencies were calculated numerically for the systems starting in very close vicinity of the Mode I of motion (the behaviour of the proper frequencies around Mode II is similar).

We note in Figure \ref{figure9} that the proper frequencies follow roughly a power law as a function of the planetary eccentricities. (This property allows us to understand the dependence on the dissipation rate described in the previous section.) The frequencies depend strongly on individual planetary masses: for small masses ($\sim 1\,M_{\rm Earth}$), the proper frequency is roughly two orders lower than for masses of sub-Saturn planets ($\sim 30\,M_{\rm Earth}$). Thus, the large deviation of the Earth-like planet path from the Mode I family observed in Figure \ref{figure12} can be explained by the effect of the aperiodic damping described in the previous section.

The passages of the system through mean-motion resonances are also observed in Figure \ref{figure12}. In Figure \ref{figure9}, they are associated with strong perturbation of the smooth evolution of the secular frequencies. It is worth noting that the perturbation by mean-motion resonances is stronger in the case of the more massive pair of planets.

\section{The orbital angular momentum transfer}\label{sec7}

In the previous sections, the study was done assuming that the orbital angular momentum of the system is conserved during  migration. Generally, this assumption is too strong and many migration mechanisms can alter the angular momentum of the system. For example, during tidal interactions, the rotation angular momentum of the central star/planet is transferred to the planet/satellite system.  During disc-planet interactions,  $\rm{AM}$ is exchanged between the planet and the disc, while, during the scattering of planetesimals by the planets, the angular momentum is removed from the system by the small bodies, etc.

Each physical process is characterized by a specific law of the $\rm{AM}$--variation. In our approach, we generalize the  process, introducing {\it ${\rm AM}$--leakage} as follows. Let $\Delta a_1$ be the change of the semi-major axis of the migrating inner planet over a small time interval $\Delta t$. After $\Delta t$, the orbital angular momentum of the system is changed by the amount $\Delta{\rm AM}_a=[{\rm AM}(a_1+\Delta a_1) - {\rm AM}(a_1)]$, where ${\rm AM}$ is also a function of the outer planet semi-major axis and eccentricities. The total variation of the orbital momentum, due to migration over $\Delta t$, is $\Delta{\rm AM}=\Delta{\rm AM}_a+\Delta{\rm AM}_e$, where $\Delta{\rm AM}_e$ is a contribution due to the change of the inner planet eccentricity.  If the angular momentum remains constant during migration of the system (i.e., $\Delta{\rm AM}=0$), the increment/decrement $\Delta{\rm AM}_a$ is absorbed by the system, according to the Equation (\ref{condition}), and we say that there is no leakage of $\textrm{AM}$ in the system.

On the other hand, some portion of $\Delta{\rm AM}_a$  may be transferred from/to the system to/from the external medium (e.g., the gaseous disc, the belt of planetesimals or the central body rotation). If $\alpha$ denotes the fraction of the loss/gain of the orbital angular momentum, the $\textrm{AM}$-variation over $\Delta t$ is
\begin{equation}\label{eq:AM-variation}
 \Delta{\rm AM}=\alpha\Delta{\rm AM}_a;\end{equation}
thus, the  condition (\ref{condition}) must be re-written as
 \begin{equation}
\Delta e^{\rm{ex}}_1 = \left(1-\alpha\right)\frac{(1-e_1^2)}{2a_1e_1}\Delta a_1.
\label{condition3}\end{equation}
Note that, defined in this form, the ${\rm AM}$--leakage can be positive or negative, depending on the value of $\alpha$. For instance, during the orbital decay of the inner planet, the system will lose the angular momentum, if $\alpha>0$, and gain it, if $\alpha<0$. Thus, varying $\alpha$, we can use the expression (\ref{condition3}) to model different laws of angular momentum exchange noted in planet-disk interactions (see Goldreich and Sari 2003, Beaug\'e et al. 2006) or in tidal interactions (see Rodr\'iguez et al. 2011a). To illustrate the dependence on $\alpha$, we assume initially that $\alpha$ is constant during the whole migration process.

\begin{figure}
\def\capfrac{1}
\centerline{\hbox{
\epsfig{figure=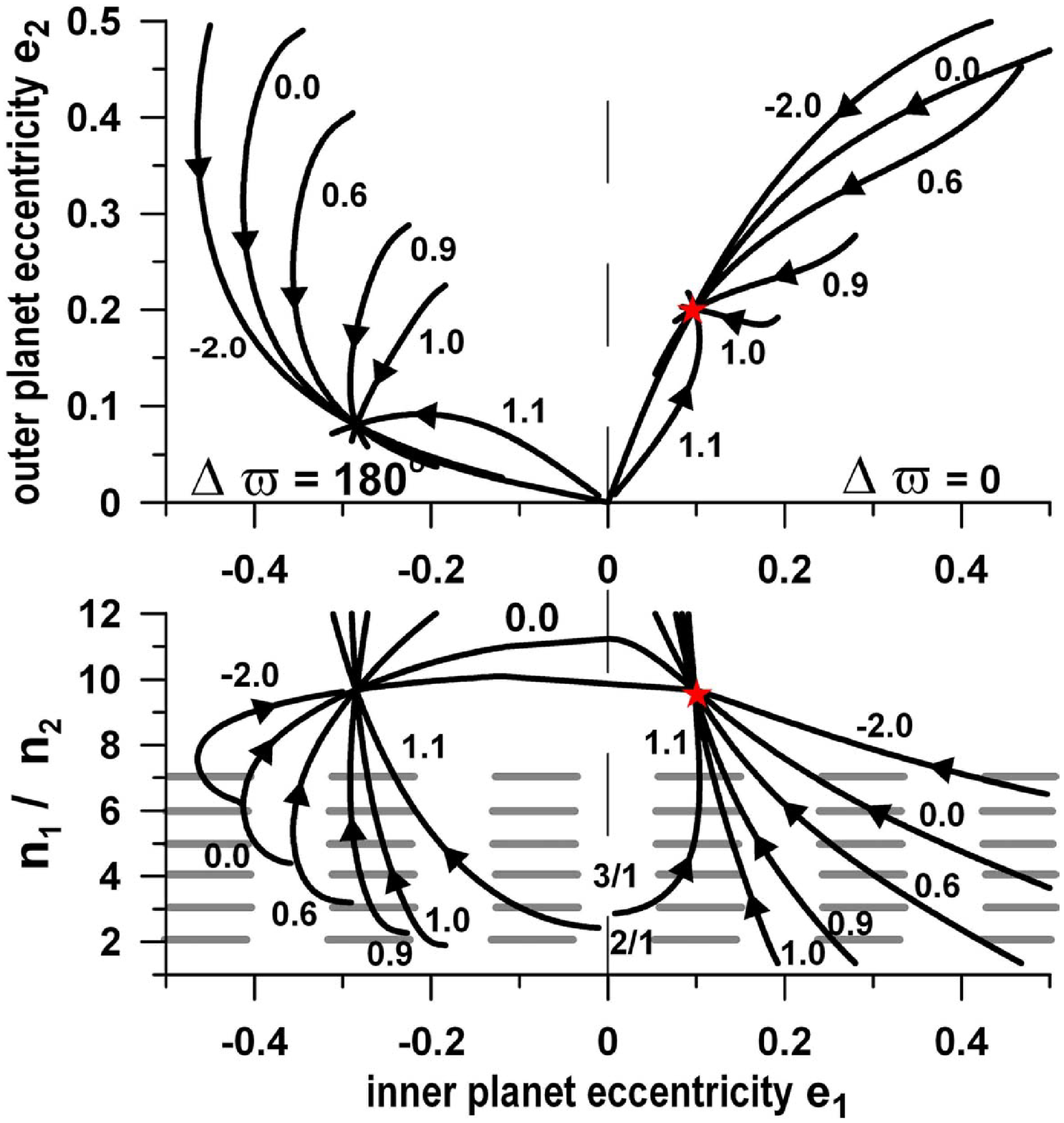, height=9cm,angle=0}
}}
\caption{Evolutionary tracks given by Mode I and Mode II on the representative planes ($e_1$,$e_2$) (top panel) and  ($e_1$,$n_1/n_2$) (bottom panel).  Both planetary masses are equal to $1.0M_{\rm J}$ and $a_2=1.0$\,AU.  The leakage of the orbital angular momentum of the system is given by the value of $\alpha$ beside each curve. Arrows show the direction of divergent orbits. The current position of the system, with $m_2/m_1=1.0$ and orbital elements $a_1=0.22$\,AU, $a_2=1.0$\,AU and $e_1=0.1$ and $e_2=0.2$ is shown by a star symbol. The routes with $\alpha > 0$ are truncated at $n_1/n_2 < 12$.
 }
\label{figure10}
\end{figure}

For $\alpha = 0$, ${\textrm{AM}}$ of the system is conserved during migration; this case was studied in the previous sections. In Figure \ref{figure10}, we show the migration route composed of both Mode I and Mode II stationary solutions, with no leakage of $\textrm{AM}$ (marked by 0), on the representative planes ($e_1$,$e_2$) and ($e_1$,$n_1/n_2$). The route was constructed for the system with the following parameters: $m_1=m_2=1.0\,M_{\rm J}$, $a_1=0.22$\,AU, $a_2=1.0$\,AU, $e_1=0.1$, $e_1=0.2$. The planets were placed on the aligned orbits, with $\Delta\varpi=0$; the position of the system on both graphs is shown by a red star.

According to Equation (\ref{eq:AM-variation}), for $\alpha > 0$, the $\textrm{AM}$ is extracted (injected) from  the system during migration when $\Delta a_1 < 0$ ($ > 0$). Four evolutionary tracks  shown in Figure \ref{figure10} were constructed with $\alpha > 0$; from these, the routes with $0<\alpha<1.0$ ($0.6$ and $0.9$) also satisfy the condition (\ref{eq:sgn}). In these cases, the divergent orbits lose the angular momentum; the system slides the corresponding routes in the direction of low eccentricities on the ($e_1$,$e_2$)--plane and the higher values of $n_1/n_2$ on the ($e_1$,$n_1/n_2$)--plane. If we assume that, to attain the current position (marked by a red star), the divergent planets acquired one of these routes, their starting eccentricities must be higher when compared to the current ones.  It is clear from Figure \ref{figure10} that, for larger values of $\alpha$, the range of the possible starting eccentricities is smaller.

For $\alpha = 1.0$,  $\Delta e^{\rm{ex}}_1 = 0$, for any $\Delta a_1$, and the whole variation of ${\rm AM}$ produced by the orbital decay is removed from the system. In this case, the dissipative force produces no change of the eccentricity, analogous to Malhotra's model (1995) of planet-planetesimal interactions. As described in Section \ref{sec2-2}, when $\Delta e^{\rm{ex}}_1 = 0$, the damped oscillations of eccentricities and the consequent trend of the system to the aligned or anti-aligned configuration do not occur. Nevertheless, as seen in Figure \ref{figure10}, the guiding route of possible migration (marked by 1.0) still exists.

In the case of $\alpha >1.0$, the variations $\Delta a_1$ and $\Delta e^{\rm{ex}}_1$ have opposite signs, according to Equation (\ref{condition3}). The decay (expansion) of the inner planet orbit is accompanied by increasing (decreasing) of eccentricities. This is because, in this case, the migration mechanism introduces an additional decrement (increment) of the angular momentum to the system, acting directly on $e_1$. One such route constructed with $\alpha = 1.1$ is shown in Figure \ref{figure10}; the direction of the evolution of the divergent system along the routes is shown by arrows. It should be emphasized that, in this case, the domains of stability of the stationary solutions are opposite to those presented in Figure \ref{figure7}.

Finally, Equation (\ref{condition3}) shows that, for $\alpha < 0$, the system gains/loses the angular momentum even in the case of the orbital decay/expansion of the inner orbit. One evolutionary track corresponding to $\alpha = -2.0$ is shown in Figure \ref{figure10}. The evolution of the system in this case is similar to the evolution along the routes with $0 \leq \alpha < 1$, but the possible starting values of the planet eccentricities would be very high.

Summarizing the information, we can say that, for $\alpha < 1.0$, the migrating system, characterized by the decreasing orbital energy  and the increasing secular one (i.e. $\Delta a_1<0$), moves in the direction of the low-eccentricity domain, with circularized orbits at its final configuration. On the contrary, when the system gains the orbital energy and loses the secular one (i.e. $\Delta a_1>0$), it moves in the direction of high eccentricity domains, where the close approaches between two planets and low-order mean-motion resonances destabilize the planetary motion. For $\alpha > 1.0$, the direction of the migrating system is inverted.

Finally, it is worth emphasizing that the migrating scenario with a constant value of $\alpha$ was assumed in this section only with the illustration propose and must be analyzed carefully for each specific migration mechanism, for instance, as has been done in (Rodr\'iguez et al. 2011b). In that paper, the authors have considered three different migration mechanisms, namely i) tidal interactions of the planetary system with the central star, ii) tidal interactions of a system of satellites with the central planet, and iii) gravitational interactions of the planetary system with the disc, modeled with a Stokes-like non-conservative force. The results obtained have shown that the ${\rm AM}$--leakage factor $\alpha$ is a function of the planet eccentricity, specific in each case. In the case of star-planet and disc-planet interactions, the typical $\alpha$--values are smaller than 1.

\section{Variation of the outer semi-major axis}\label{sec8}

Some migration processes can be described assuming that the dissipation affects only the outer planet orbit. For instance, this assumption is in good accordance with hydrodynamical simulations, which show the outer planet driven inward due to torques exerted on it by outside nebular material (Kley 2000, 2003). Our model can be easily adopted to this case: The possible evolutionary tracks which the migrating system could take to attain its current configuration, are obtained through the calculation of Mode I and Mode II solutions of the Hamiltonian (\ref{eq:Hamilnumeric}), for all possible values of $a_2$ and with a fixed $a_1$.

\begin{figure}
\def\capfrac{1}
\centerline{\hbox{
\epsfig{figure=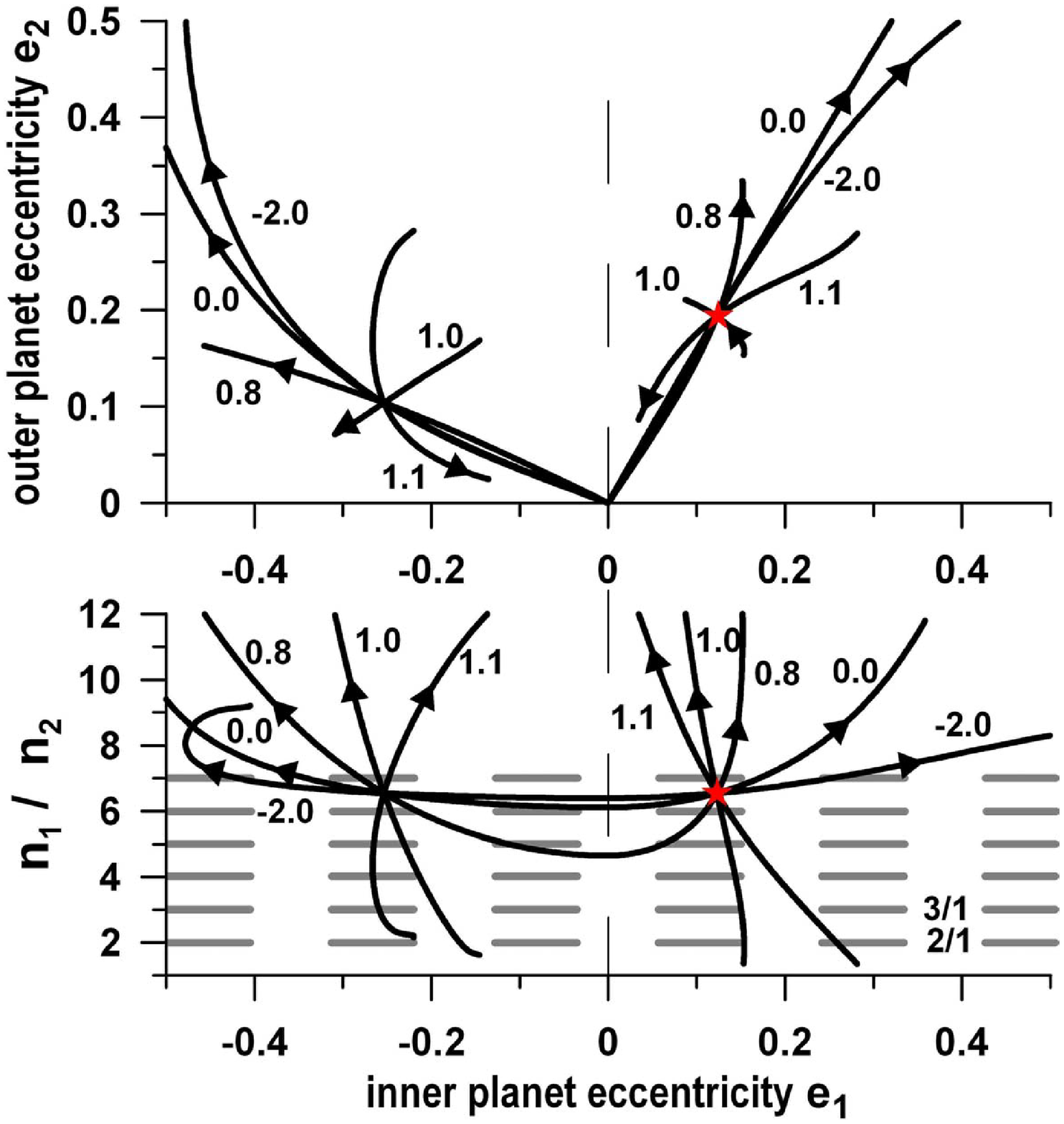, height=9cm,angle=0}
}}
\caption{Same as in Figure \ref{figure10}, except the families were calculated assuming all possible values of $a_2$ and a fixed $a_1=0.02$\,AU.
 }
\label{figure13}
\end{figure}

Figure \ref{figure13} shows by a star symbol the current position of a hypothetical two-planet system defined by the planet masses $m_1=m_2=1.0\,M_{\rm Jup}$, the semi-major axes $a_1=0.2$ and $a_2=0.7$, the eccentricities $e_1=0.12$ and $e_2=0.2$, and the aligned periastra, $\Delta\varpi=0$. In analogy with what has been done in the case of the migrating inner planet, we calculate the evolutionary tracks parameterized by several values of the parameter $\alpha$; they are present in Figure \ref{figure13}. Recall that $\alpha$ characterizes the ${\rm AM}$--leakage in the system, when  $\alpha=0$ corresponds to the case of the conservation of the orbital angular momentum of the system during migration. To model the ${\rm AM}$--leakage in the system with the migrating outer planet, we used the Equation (\ref{condition}), with the index $1$ replaced by the index $2$.

Figure \ref{figure13}, when compared to Figure \ref{figure10}, shows that the evolutionary routes of the migrating system are similar in both cases. Moreover, the domains of stability of the secular Modes, I and II, which determines the mode which will guide the migrating system, are the same (see Figure \ref{figure7}). The main difference is that divergent orbits are now associated with increasing eccentricities, if the increase of the orbital angular momentum due to increase of $a_2$ is not lost totally ($\alpha < 1.0$), according to the condition (\ref{condition3}) with the index $1$ replaced by $2$.  Therefore, the migrating systems characterized by the simultaneously increasing orbital and secular energy, move in the direction of the high eccentricity domains, where the close approaches between two planets could originate a large-scale instability.  The situation is inverted for convergent orbits, when the systems lose both the orbital and secular energy and, for $\alpha < 1.0$,  move in the direction of the low-eccentricity domain, with totally circularized orbits at their final configurations.

\begin{figure}
\def\capfrac{1}
\centerline{\hbox{
\epsfig{figure=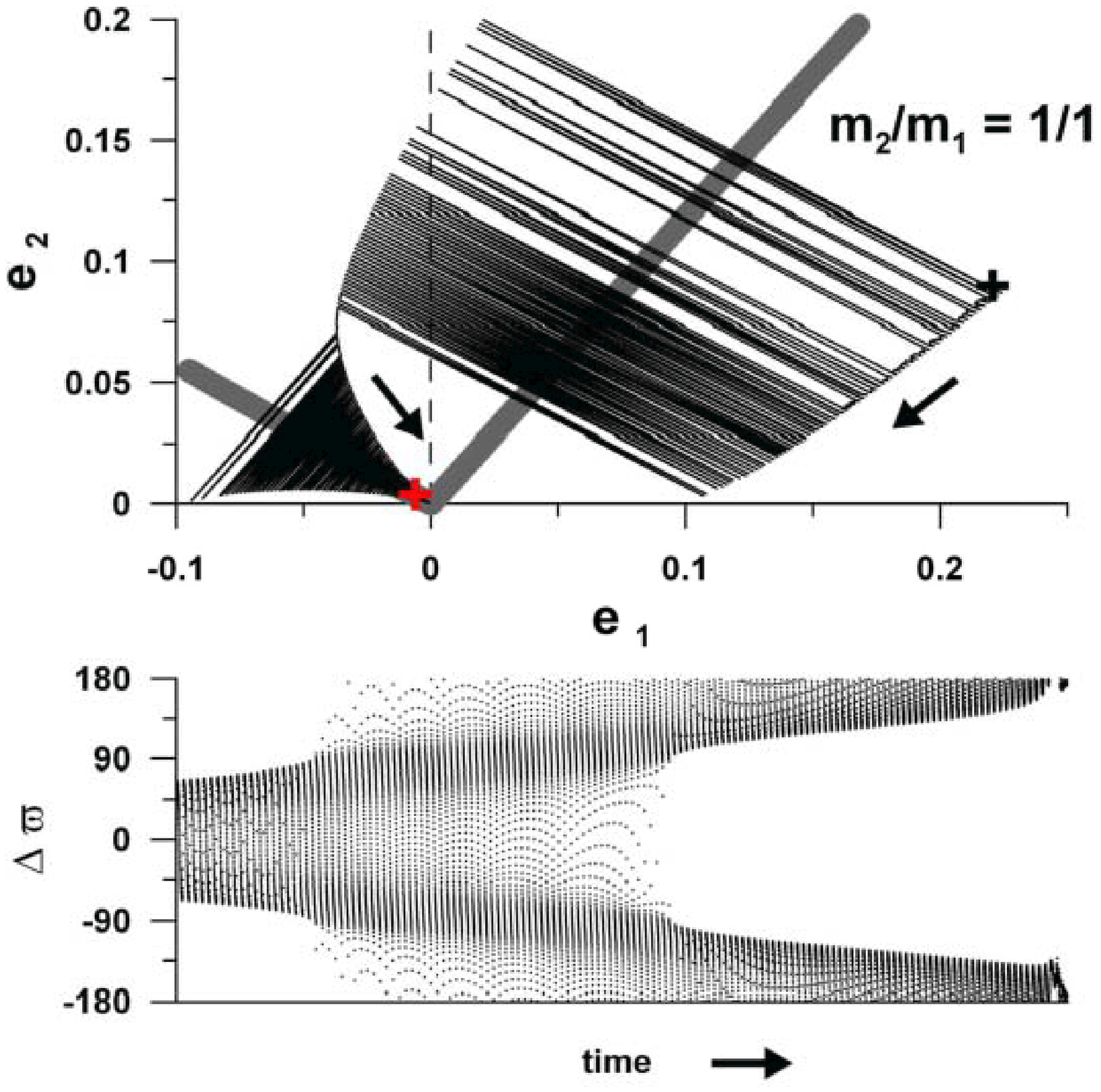, height=9cm,angle=0}
}}
\caption{Numerical simulation of the migrating two-planet system. Top: The starting position of the system is shown by a black symbol, while its final position by a red symbol. Families of stationary solutions are plotted by thick gray line.  Arrows show the direction of the evolution of the system. Bottom: Time evolution of the secular angle $\Delta\varpi$.
 }
\label{figure14}
\end{figure}

We simulated the migration of a hypothetical system, assuming that the outer planet experienced inward migration losing its orbital energy. For the sake of simplicity, we assume that, during migration, the orbital angular momentum of the system is conserved, i.e., $\alpha = 0$. The obtained migration path is shown in Figure \ref{figure14}. The system with $m_1=m_2=1.0\,M_{\rm Jup}$ started at $a_1=0.02$\,AU and $a_2=0.049$\,AU ($n_1/n_2=3.83$), oscillating around $\Delta\varpi=0$; the starting position is shown by a black symbol in Figure \ref{figure14}. Evolving from the oscillation to the circulation regime, the system was converging gradually to the anti-aligned configuration (Mode II of motion) and finished its evolution on the totally circularized orbits of the planet, at $n_1/n_2=3.6$. The time evolution of the secular angle $\Delta\varpi$ shown on the bottom panel in Figure \ref{figure14}, illustrates how the system, which was initially oscillating around $0$, was smoothly evolving into the Mode II of motion with small oscillations around $180^\circ$. Given that $\sqrt{a_1/a_2} < m_2/m_1$ during the evolution the system, this results is in good agreement with our model (see Figure \ref{figure7}).

\section{Discussions}\label{sec9}

In this paper we have presented a simple method for a qualitative study of the secular dynamics of the two-planet migrating system. Our model is valid to describe the evolution of the planets which are orbiting on the same plane, avoiding any mean-motion resonance, and are not too close to produce strong short-period interactions between them.
The evolution of the planets was studied from the point of view of the variation of integrals of motion of the conservative secular problem: energy, orbital angular momentum and planetary semi-major axes. In this formulation, no particular dissipative mechanism was introduced; therefore, the results obtained are valid for any migration process. As a consequence, the detailed knowledge of the physical parameters of the migration process and starting configurations of the system are unneeded in this qualitative analysis. The only assumption need to be done is that the dissipation process is weak and slow sufficiently.

We have shown that, under this assumption, the evolutionary routes of migrating planets in phase space are traced by stationary solutions of the conservative secular problem. Parameterized by mass ratios, the families of the stationary solutions were constructed for continuous values of the semi-major axes ratio and fixed values of the angular momentum of the system, assuming initially that it is conserved during migration. We have shown that each family is composed of two distinct branches, corresponding to the Mode I and Mode II solutions, which are connected only at the origin.  During migration, the system is attracted to and follows one of these branches, which is said to be stable. In the case of divergent orbits, the Mode I (aligned orbits) is stable when the mass ratio and the instantaneous semi-major axes ratio satisfy the condition $\sqrt{a_1/a_2} < m_2/m_1$; in the opposite case, the Mode II (anti-aligned orbits) is stable. For convergent orbits, the stability of the secular modes is inverted.

We have further shown that the effects of the gain/loss of the orbital angular momentum of the system during migration modifies the evolutionary tracks in the phase space. Knowing that the angular momentum exchange is specific for each kind of dissipative forces, we generalize our study, introducing $\textrm{AM}$-leakage. The $\textrm{AM}$-leakage is defined through the factor $\alpha$ (see Equation (\ref{eq:AM-variation})), where  $\alpha$ is a fraction of the angular momentum variation produced by migration (i.e. by the variations of $\Delta a_{1}$ and/or $\Delta a_{2}$), which is extracted (or added) from the system during migration. When there is no $\textrm{AM}$-leakage in the migrating system (i.e. $\alpha=0$), the variation of $\textrm{AM}$ due to migration is totally absorbed by the system, in such a way that its orbital angular momentum is conserved. In this case, the eccentricity of the migrating planet is affected according to Equation (\ref{condition}). The case of $\alpha=1$ corresponds to loss/gain of the total amount of the $\textrm{AM}$-variation produced by migration and, in this case, the eccentricities of the planets are not affected by external forces. The values of $\alpha$ greater then 1 are characteristic of processes which produce trends of the semi-major axis and the eccentricity with opposite signs. It is worth noting that the range of theoretical values of $\alpha$ is large; however, the analysis of the orbital angular momentum exchange in several migrating systems driven by specified non-conservative forces (Rodr\'iguez at al. 2011b), has shown that, in practice, the typical values of $\alpha$ belong to the interval below 1.

\begin{figure}
\def\capfrac{1}
\centerline{\hbox{
\epsfig{figure=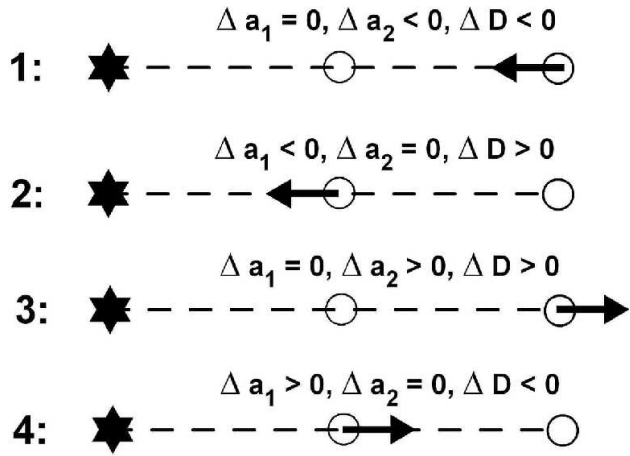, height=6cm,angle=0}
}}
\caption{Schematic view of possible one-planet migration scenarios in two-planet systems. }
\label{figure15}
\end{figure}

Analyzing  the evolutionary routes in the phase space we can i) conjecture on the past evolution of the system to its present location in the phase space; and ii) do robust predictions about the possible final configurations of the currently evolving system. We have shown that the evolution of the secular system, under weak dissipative forces, and its final state depend on the balance between the orbital and secular energy variations of the system. To clearly illustrate this feature, we decompose possible migration processes into four elementary components; they are schematically shown in Figure \ref{figure15}.

The scenario \textbf{1} describes approximately the evolution of the system interacting with the outer protoplanetary disc. The planet-disc interactions are frequently modeled using the non-conservative Stokes force, which is characterized by the $\textrm{AM}$-leakage factor $\alpha < 1$ (Rodr\'iguez et al. 20011b). During migration, the system loses both the orbital and secular energy; as a consequence, both orbits are circularized during the secular evolution, while their mutual distance decreases, allowing a smooth entrance and a nearly zero-amplitude evolution inside a low-order mean-motion resonance, for instance, the 2/1 resonance. Actually, this is the most populated resonance, with the five known planet pairs evolving inside it: GJ876 \textbf{c}-\textbf{b}, HD40307 \textbf{c}-\textbf{d}, HD73526 \textbf{b}-\textbf{c}, HD82943 \textbf{c}-\textbf{b} and HD128311 \textbf{b}-\textbf{c}. Analyzing these systems, we have found that their parameters systematically satisfy the condition $\sqrt{a_1/a_2} < m_2/m_1$. Thus, if these systems evolved according to scenario \textbf{1}, they would converge to the Mode II of the secular motion and approach the 2/1  resonance in the anti-aligned configurations on low-eccentricity orbits. As shown in (Michtchenko et al. 2008\,a,b), this configuration is the only way to enter smoothly into the strong mean-motion 2/1 resonance. Several numerical simulations performed (Lee and Peale 2002, Ferraz-Mello et al. 2003, Beaug\'e et al. 2006, Hadjidemetriou and Voyatzis 2010, among many others) confirm the described behaviour of migrating systems.

The case \textbf{2} describes the orbital decay of a close-in planet due to tidal interactions with the slowly rotating central star. In this case, the $\textrm{AM}$-leakage factor $\alpha < 1$ (Rodr\'iguez et al. 2011b) and the system loses the orbital energy and gains the secular energy; as a consequence, both orbits are totally circularized, while their mutual distance is increased. The planets \textbf{b} and \textbf{c} of the CoRoT-7 system have probably reached this configuration (Ferraz-Mello et al. 2011). The investigations of the behaviour of the hypothetical systems  HD 83443 (Wu and Goldreich 2002) and  Gliese 436 (Batygin at al. 2009) have provided the results in complete agreement with our model.

An interesting example is found in our Solar System: the coupled motion of the longitudes of peristra of the pair Jupiter-Uranus, whose difference $\Delta\varpi=\varpi_{\rm{U}}-\varpi_{\rm{J}}$ oscillates around $180^\circ$ with an amplitude about $70^\circ$ degrees (Milani and Nobili 1984). We can show that this peculiar configuration of the outer planets could contain a record of planet migration in the past. Indeed, Fern\'andez and Ip (1984, 1996) have shown that, during the scattering of planetesimals in the final stage of the Solar System formation, the giant planets experienced the migration evolution: three outer giant planets experienced an outward displacement, while Jupiter, as the innermost giant planet and main ejector of bodies, migrated sunward. As shown in Milani and Nobili (1984), the outer Solar System can be represented by two main subsystems, Sun-Jupiter-Saturn and Sun-Uranus-Neptune, exchanging the angular momentum over a period of 1.1\,Myr, same of the $\Delta\varpi$--oscillation. In this three-body approximation, the orbits of two subsystems were divergent during migration and the whole system gained the secular energy. In the meantime, the system lost its orbital energy, due to the dominant mass of Jupiter. This scenario corresponds to the scheme  \textbf{2} in Figure \ref{figure15}, when the divergent system, whose parameters satisfy the condition $\sqrt{a_1/a_2} > m_2/m_1$, is captured in the Mode II of motion, oscillating around $180^\circ$.

There are not many examples of the planetary systems to illustrate the migration scenario \textbf{3}, when the system gains both the orbital and secular energy. In this case, the mutual planetary distance continuously increases, together with the eccentricities of the planetary orbits (under condition $\alpha < 1$). Possibly, the exosystems characterized by large mutual distances, sometimes referred to as hierarchical systems, have experienced this kind of evolution in the past, and this fact could explain consistently high eccentricities of the planetary orbits in such systems.  On the other hand, the orbits of the hierarchical planet pairs do not exhibit the capture inside of one of the secular modes of motion characterized by oscillations of the secular angles around $0$ or $180^\circ$; thus, this  point  needs  to  be further  addressed in  more detail.

Finally, during migration described by the scenario \textbf{4} in Figure \ref{figure15}, the system gains the orbital energy and loses the secular one. For $\alpha < 1$, the eccentricities of two convergent orbits continuously increase during migration and the secular system is ultimately disrupted when it approaches a low-order mean-motion resonance. Thus, it seems to be highly improbable to find real systems which have experienced the large-scale evolution  corresponding to the case \textbf{4}, at least for the values  of the $\textrm{AM}$-leakage factor $\alpha < 1$. The picture is different in the case of $\alpha > 1$, which is characteristic for tidal migrations in planet-satellite systems, specially those formed by giant planets  (see Rodr\'iguez at al. 2011b). In this case, the convergent orbits will be accompanied by decreasing eccentricities, similar to what occurs in the case \textbf{1}.

As a final consideration, we discuss the advantages and disadvantages of the method introduced in this paper. The main advantage, in our opinion, is that the method allows us to study the secular motion of the system evolving under a generic mechanism of migration. Thus, the patterns of the secular behaviour of the migrating system obtained in such a way, are universal. The model provides a robust prediction of the possible starting and final configurations of the systems undergone to migration processes, despite the qualitative approach used in its construction. The principal restriction of the model comes from the same fact that an unspecified type of dissipative forces was invoked for the dynamical study. Indeed, each migration process is described by specific laws of the time variations of the energy, the orbital angular momentum and the semi-major axes of the planets. Treating the problem in general terms of variation of these quantities, we lose the information on the migration time-scales, characteristic of the system under study.
Thus, to obtain a complete solution of the problem, our model must be completed by additional analytical (e.g. Mardling 2007) or numerical (e.g. Rodr\'iguez et al. 2011) investigations of the time-dependence of the secular evolution on a specific dissipative force.

\begin{appendix}

\section{The basic concepts of the conservative secular dynamics }\label{sec1}

\subsection{Secular model}\label{sec1-1}

In order to understand the secular behaviour of the planet pair, we study the second-order (i.e., ${\cal O}(e^2)$) approximation of the Hamiltonian describing the secular planar three-body system reduced to one degree of freedom (for details see Michtchenko \& Ferraz-Mello 2001, Callegari et al. 2004). Discarding constant Keplerian terms, the reduced secular Hamiltonian can be written, in terms of the canonical elliptic variables (\ref{eq:1-3}),  as
\begin{equation}\label{eq:hamil}
{\mathcal H}_{\rm Sec} =  2(C-D)\,I_{1} +2 E\sqrt{I_{1}({\rm AMD}-I_{1})} \cos\Delta\varpi,
\end{equation}
where ${\rm AMD}$ is given in (\ref{eq:AMD}) and the coefficients are defined as
\begin{equation}
\begin{array} {lll}
C &=& -(2)^{(0)}\,M_3/{L_1 L_2^2},\\
D &=& -(3)^{(0)}\,M_3/L_2^3,\\
E &=& -(21)^{(-1)}\,M_3/{L_2^2\sqrt{L_1\,L_2}}\,.
\end{array}\label{coef}
\end{equation}
The coefficients $(2)^{(0)}=(3)^{(0)}$ and $(21)^{(-1)}$ are functions of the Laplace coefficients $b_s^{(k)}(a_1/a_2)$ written in the form adopted by Le Verrier (1855)  (for their explicit expressions, see Callegari et al. 2004) and
$M_3=\mu_2^2\,m_1\,m_2^{\prime 3}$.

The long-term dynamics of two massive planets evolving far from any mean-motion resonance is defined by the averaged Hamiltonian (\ref{eq:hamil}); in the non-dissipative approach, ${\mathcal H}_{\rm Sec}$ is the secular energy of the system, which is conserved along the motion. As the energy ${\mathcal H}_{\rm Sec}$, the orbital angular momentum (\ref{eq:AM}) and the angular momentum deficit (\ref{eq:AMD}) are also constants of the  of the secular planet motion (they were used to reduce the usual two-degrees-of-freedom Hamiltonian).

The ${\rm AMD}$ appearing as a parameter in the expression of the secular Hamiltonian (\ref{eq:hamil})  has a clear algebraical interpretation:  it has a minimum value (zero) for circular orbits and increases with increasing eccentricities. The behaviour of ${\rm AM}$ is reverse of that of ${\rm AMD}$.
It should be emphasized that the ${\rm AMD}$ is conserved only in the secular problem because, in this problem, $L_1$ and $L_2$ (i.e., $a_1$ and $a_2$) are constants of motion. Also, the action variable  $I_1$  and its analogous $I_2$ given in (\ref{eq:1-3}) define  the partial angular momentum deficits of the planets (their sum  is the  ${\rm AMD}$).

\subsection{Phase space}\label{sec1-2}

In order to have a geometrical representation of the phase space of the secular system, we transform the Hamiltonian (\ref{eq:hamil}) following the approach introduced by Pauwels (1983). First, we note that the coefficient $E$ given in (\ref{coef}) is always positive, while $C$ and $D$ are always negative. We introduce a new variable $\delta$ and new parameters  $\alpha$ and $\beta$, in the following way:
\begin{equation}
{\mathcal H}_{\rm{Sec}} = 2 \overbrace{(C-D)}^{\beta\cos \alpha}\underbrace{I_{1}}_
{ \frac{{\rm AMD}}{2}(1+\cos \delta)}
 + 2\overbrace{E}^{\beta\sin \alpha}\underbrace{\sqrt{I_{1}\,({\rm AMD}-I_{1})}}_{\frac{{\rm AMD}}{2}\sin \delta} \cos\Delta\varpi,
\label{eq2}\end{equation}
where both $\alpha$ and $\delta$ vary in the range from $0$ to $\pi$. We also have $\beta^2 =(C-D)^2+E^2$ and Equation (\ref{eq2}) then is rewritten as
\begin{equation}\label{eq3}
\,{\mathcal H}_{\rm{Sec}}-{\beta\,\rm AMD}\cos \alpha = {\beta\,\rm AMD}(\cos \alpha\cos \delta + \sin \alpha\sin \delta\cos\Delta\varpi).
\end{equation}

\begin{figure}
\centerline{\hbox{
\epsfig{figure=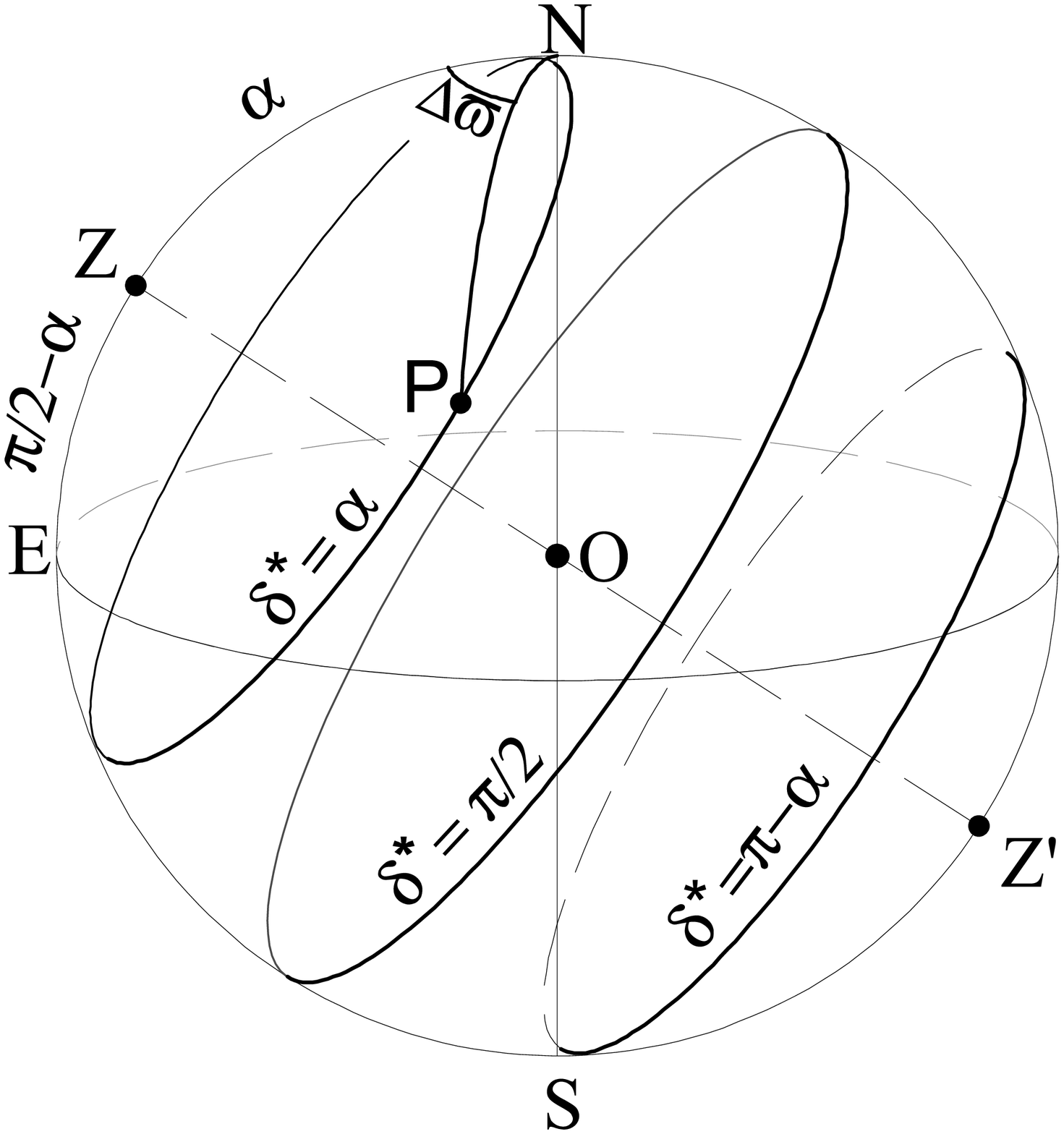, height=4.3cm,angle=0}\hspace*{0.8cm}
\epsfig{figure=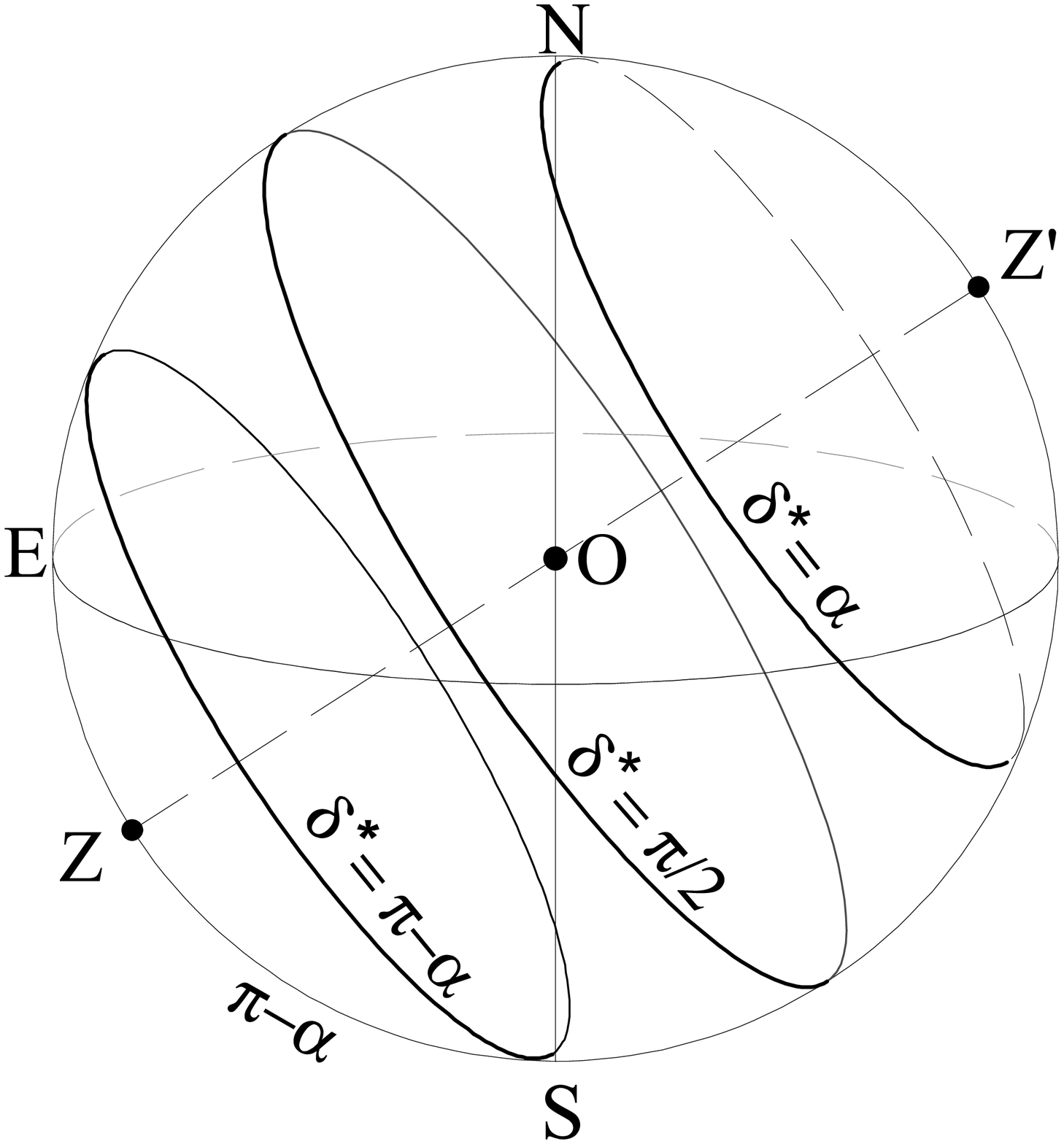, height=4.3cm,angle=0}
}}
\caption{Separation of the $\Delta\varpi$ modes of motion  on Pauwels' sphere. In the spherical caps defined by parallel circles passing through the N and S poles of the sphere, $\Delta\varpi$ oscillates: around 0 in the cap with $Z$ vertex and around $\pi$ in the cap with $Z^\prime$ vertex. In the zone between the two parallels, $\Delta\varpi$ circulates. {\it Left}: Case $0<\alpha<\pi/2 $. {\it Right}: Case $\pi/2 < \alpha < \pi$.}
\label{bolas}
\end{figure}

The variables ($\delta$, $\Delta\varpi$) may be considered as defining a system of spherical coordinates, with a vertical polar axis $NS$, on a sphere (Pauwels' sphere) of radius ${\beta\,\rm AMD}$.  The right-hand side of Equation (\ref{eq3}) is the expression of a cosine law and can be re-written as
\begin{equation}\label{eq3b}
{\mathcal E}_1 = \,{\mathcal H}_{\rm{Sec}} - {\beta\,\rm AMD}\cos \alpha = {\beta\,\rm AMD}\cos  \delta^*\, ,
\end{equation}
where $\delta^*$ is the distance of the generic point P with spherical coordinates ($\delta$, $\Delta\varpi$) to the pole $Z$, defined by the axis $Z\,Z^\prime $, which is inclined of $\alpha$ with respect to the vertical $N\,S$ (see Figure \ref{bolas}). The points with $\delta^*={\rm const}$ form parallel circles on the sphere, whose centers belong to the axis $Z\,Z^\prime$. The intersections of the axis $Z\,Z^\prime$ with the sphere have spherical coordinates $\delta=\alpha$ and $\Delta\varpi=0$, for $Z$--pole; for $Z^\prime$--pole, $\delta=\pi-\alpha$ and $\Delta\varpi=\pi$. These intersections correspond to the two equilibria of the one-degree-of-freedom secular Hamiltonian (\ref{eq3b}).

$\delta^*$ is the polar angle in a system of spherical coordinates whose axis is $Z\,Z^\prime$ (instead of the vertical axis).  In analogy with the transformation given in Equation  (\ref{eq2}), we may introduce a new variable  $I^*_{1}$ through
\begin{eqnarray}\label{eq:I1}
{I^*_{1}}&=&{\frac{{\rm AMD}}{2}(1+\cos \delta^*)}
\label{eq:I2}\end{eqnarray}
and write the secular Hamiltonian (\ref{eq3b}) as
\begin{equation}\label{eq3c}
{\mathcal E}_1 = \beta\,(2I_1^*-{\rm AMD}).
\end{equation}

Some important consequences of the performed transformations come from the fact that $I^*_1$ may be considered as a new {\it action variable}. The conjugate of $I^*_1$ is an angle $w_1$, rotating with the constant angular velocity $\dot{w}_1= 2\beta$, which is independent on the energy. As a consequence, the periods of the motions on all parallel circles are same and depend only on $\beta$, i.e., on the constants $C$, $D$, $E$.

\begin{figure}
\centerline{\hbox{
\epsfig{figure=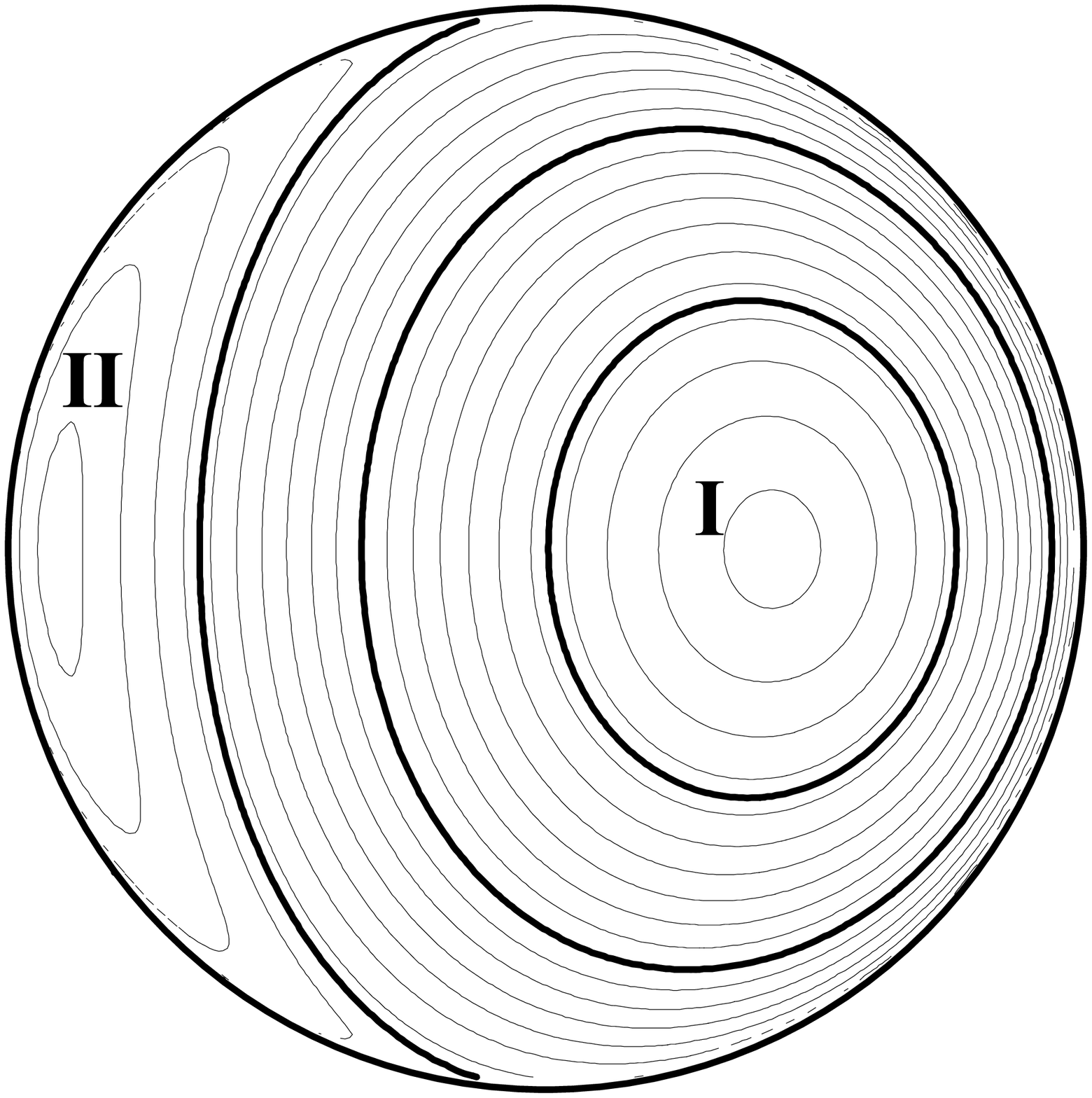, height=4cm,angle=0}\hspace*{0.8cm}
\epsfig{figure=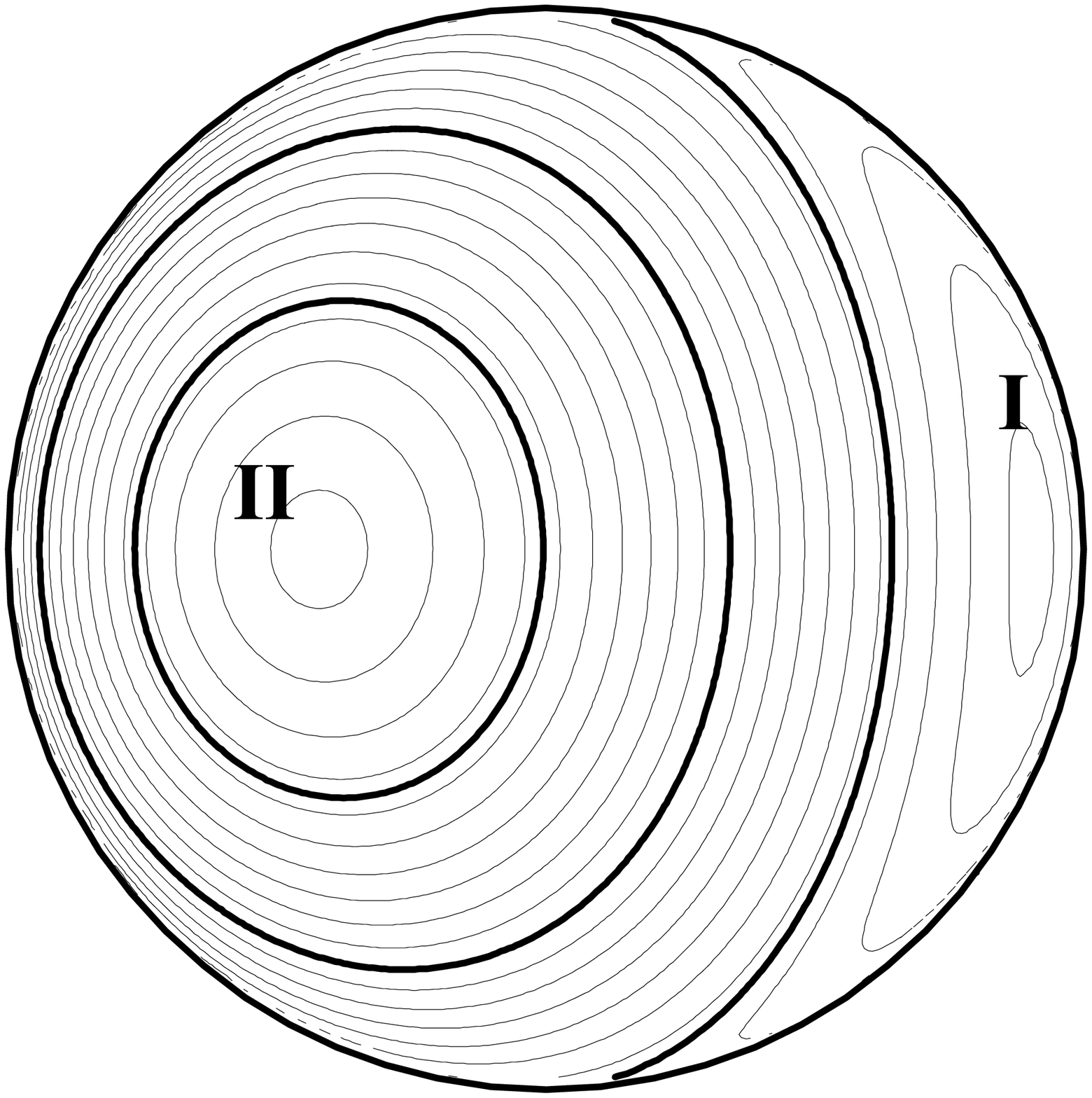, height=4cm,angle=0}
}}
\caption{Locus of the curves $I_1^* = {\rm const}$ in the plane $I_1 \cos\Delta\varpi, I_1 \sin\Delta\varpi$. The parallel circles shown in Figure \ref{bolas} appear here as continuous lines. In both cases, $I_1^*$ is maximum at the Mode I center. {\it Left}: Case $0<\alpha<\pi/2 $. {\it Right}: Case $\pi/2 < \alpha < \pi$. }
\label{planos}
\end{figure}

The new action variable $I^*_{1}$ (as well as the polar angle $\delta^*$) is a constant of the motion. Since $\beta AMD > 0$, the secular energy ${\mathcal E}_1$ and, therefore, ${\mathcal H^*}_{\rm{Sec}}$ is  maximal  (resp. minimal ) at the pole $Z$, where $\delta^*=0$ (resp.  $Z^\prime$, where $\delta^*=\pi $).

The dynamical interpretation of the solutions defined by the secular Hamiltonian is now simple. For a given AMD, the solutions are the curves on the sphere corresponding to points of the same energy, that is, parallel circles. Indeed, the conservation of the energy and of the orbital angular momentum (or  the angular momentum deficit) along one solution  define the main features of the secular motion of planetary systems.

The solutions in the spherical zone between the two parallel circles, defined by $\delta^* = \alpha $ and $\delta^* = \pi-\alpha $ (see Figure \ref{bolas}), circulate around the vertical axis $N\,S$ and, along them, the secular angle $\Delta\varpi$ circulate from 0 to $2\pi$. In the solutions inside the spherical caps between these parallels and the poles $Z$ and $Z^\prime$, the angles $\Delta\varpi$ remain bounded. In the spherical cap whose vertex is at $Z$ (resp. $Z^\prime$), the angle $\Delta\varpi$ remains in the interval $-\pi/2<\Delta\varpi<\pi/2$ (resp. $\pi/2<\Delta\varpi<3\pi/2$) the perihelia move in such a way that $\Delta\varpi$ oscillates around 0 (resp. $\pi$). It is worth noting that this behaviour does not depend on which hemisphere on the Pauwels' sphere each pole is located.

For the negative coefficients $C$ and $D$, when $|C| < |D|$ (resp. $|C| > |D|$), $\alpha < \pi/2$ (resp. $\alpha > \pi/2$) and the pole $Z$ appears in the upper (resp. lower) hemisphere of the Pauwels' sphere (see Figure \ref{bolas}). The transition between the two hemispheres occurs when $C=D$ (or $L_1=L_2$) and, in this particular case, both centers are located on the horizontal plane of the Pauwels' sphere.  This last condition can be rewritten, up to second order in masses, as in (\ref{condition1}).

Finally, let it be said that the derivations given above can be done using the variables ($-I_2$, $\Delta\varpi$). Indeed, it is sometimes useful to use simultaneously the two possibilities for better characterizing the planets dynamics. The results are equivalent to the above ones and are obtained just putting $I_1={\rm AMD}-I_2$ in the equations.

\subsection{Stationary solutions}\label{sec1-3}

As shown in Figures \ref{bolas} and \ref{planos}, two possible mutual orientations of the planet orbits are of special interest to understand the secular dynamics: one is the case of aligned orbits, when $\Delta\varpi = 0$, and the other is the case of anti-aligned orbits, when $\Delta\varpi = 180$. These particular configurations correspond to the stationary solutions of the secular three-body (two planet) problem and have been dubbed as  Mode I and Mode II (Tittemore and Wisdom 1988, Michtchenko and Ferraz-Mello 2001).

The Mode I corresponds to the center located at the pole $Z$ on the Pauwels' sphere, while the Mode II corresponds to the center located at the pole $Z^\prime$. It is worth mentioning that, by the definition (\ref{eq:I1}), $I_1^*$ is equal to zero (minimum of $I_1^*$) at the center II and equal to ${\rm AMD}$ (maximum of $I_1^*$) at the center I.
In the alternative representation, when $I_2$ is used instead of $I_1$, the picture is similar to Figure \ref{planos},  just by commuting the left and right panels. We introduce $I_2^*={\rm AMD}- I_1^*$ and then we have that $I_2^*$ is equal to zero (minimum of $I_2^*$) at the center I and equal to ${\rm AMD}$ (maximum of $I_2^*$) at the center II.

\subsection{The general secular problem}\label{sec1-4}

It is worth stressing the fact that the secular problem defined by the Hamiltonian (\ref{eq:hamil}) is a low-order approximation valid only for small eccentricities and large separations between planets. To expand the model to high eccentricities, we use a semi-analytical approach employing a numerical averaging of the short-period gravitational interactions of the two planets (Michtchenko and Malhotra 2004). The Hamiltonian describing precisely the secular perturbations between the planets on high-eccentricity orbits, is given by
\begin{equation}
{\mathcal H}_{\rm Sec} = -\frac{1}{{(2\pi)^2}}\int_0^{2\pi}\int_0^{2\pi}\frac{G\,m_1\,m_2}{a_2} \, R(a_i,e_i,M_i,\varpi_i)\,dM_1dM_2,
\label{eq:Hamilnumeric1}
\end{equation}
where the integrand is the direct part of the disturbing function written in terms of the canonical astrocentric semi-major axes and eccentricities of the planets ($a_i$ and $e_i$, respectively) and of the angular elements: mean anomalies $M_i$ and longitudes of pericenter $\varpi_i$.

The phase space is no longer the same as shown in Figures \ref{bolas} and \ref{planos}, since those structures are now perturbed by the high-order eccentricity terms. The role played by the cases $|C| < |D|$ and $|C| > |D|$ is now more involving. However, the important characteristics continues to be the fact that the pole Z (associated with the maximum of the secular energy) is located either in the upper or in the lower hemisphere. The transition between two cases occurs when the pole Z appears on the horizontal plane of the Pauwels' sphere, i.e., when $\delta_Z=\pi/2$ or $I_1^Z=\frac{1}{2} {\rm AMD}$. We may still write this condition as $I_1^Z=I_2^Z$ or, up to second order in masses, as:
\begin{equation}
\frac{m_1}{m_2}\,\sqrt{\frac{a_1}{a_2}}= \frac{1-\sqrt{1-e^{Z\,2}_2}}{1-\sqrt{1-e^{Z\,2}_1}},
\label{condition2} \end{equation}
where $e_1^Z$ and $e_2^Z$ are the eccentricities of the stationary solutions. The linear approximation (\ref{condition1}) results from the above expression when $e_1^Z=e_2^Z$. It should be stressed, that the condition (\ref{condition2}) defines the mode of motion to which the secular system evolves when dissipative effects are included.

\begin{figure}
\def\capfrac{1}
\centerline{\hbox{
\epsfig{figure=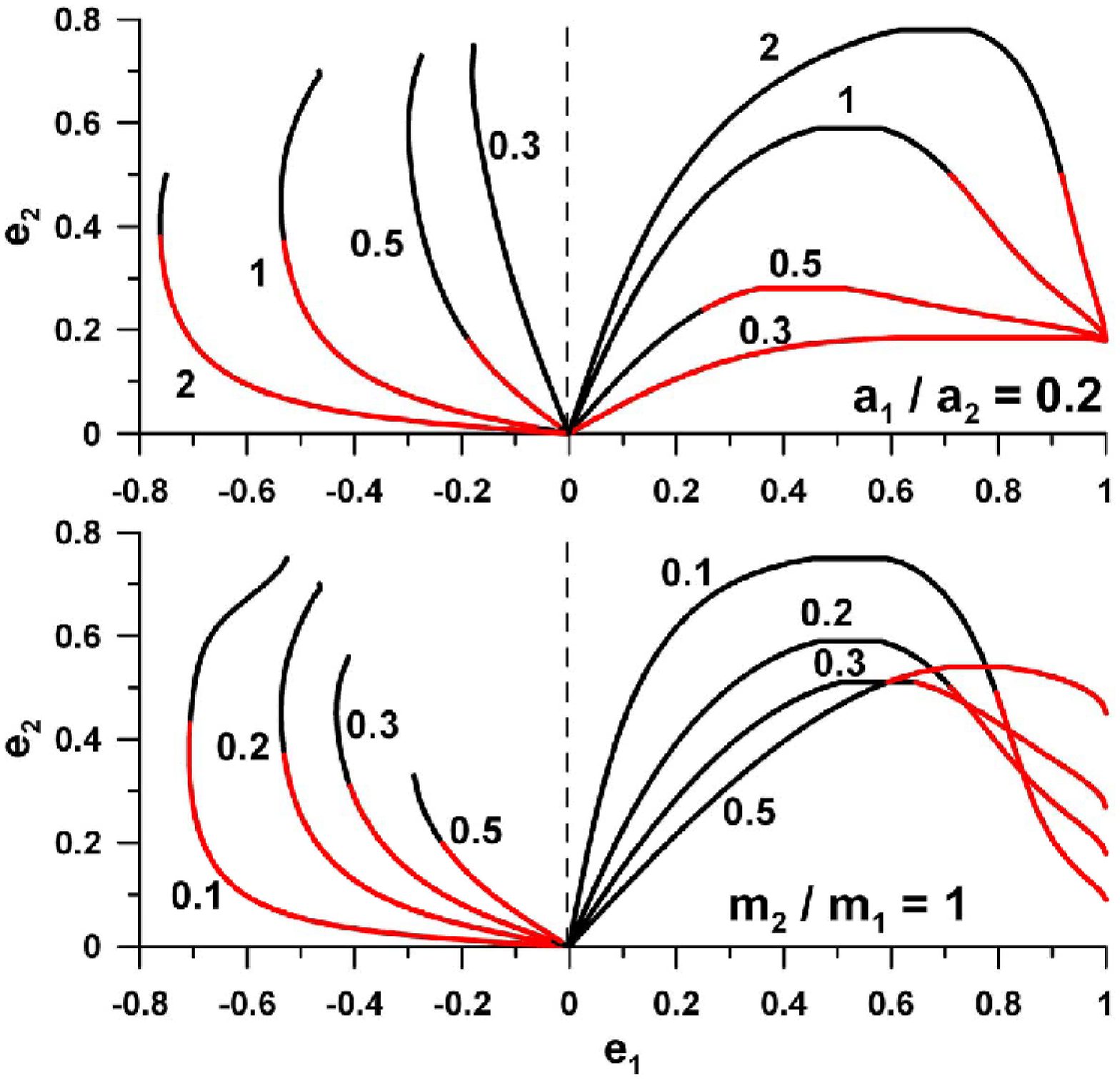, height=8cm,angle=0}
}}
\caption{Families of stationary solutions of the conservative secular problem, parameterized by the planet mass ratio (top panel) and the semi-major axes ratio (bottom panel). Mode I solutions ($\Delta\varpi=0$) are in the right half-plane, with positive values on the e1-axis. Mode II solutions ($\Delta\varpi=180^\circ$) are in the left half-plane, with negative values on the e1-axis. The branches of the curves in black color corresponds to solutions with $I_1^Z < I_2^Z$ and in red color to solutions with $I_1^Z > I_2^Z$. Along each curve, the values of the parameters $m_2/m_1$ and $a_1/a_2$ are fixed, but the values of the energy and the angular momentum vary.}
\label{figure3}\end{figure}

Figure \ref{figure3} shows the families of stationary solutions parameterized by the ratio of the planet masses and by the semi-major axes ratio. In the top panel, the ratio of the semi-major axes is fixed at $a_1/a_2=0.2$; in the bottom panel, the mass  ratio is fixed at $m_2/m_1=1$. Along each curve, the values of both parameters are fixed, but the values of the energy and the angular momentum deficit vary. We calculated the values of the partial angular momentum deficit of the planets for each stationary solution shown in Figure \ref{figure3}. The relation between them varies along each curve: when $I_1^Z < I_2^Z$, we plot the branches of the families in black color, and, when $I_1^Z > I_2^Z$, we plot them in red color.

\section{Elements of the dissipative secular dynamics}\label{sec2}

In this section we illustrate the main features of the dynamics of secular systems with dissipation. For this task we use the simple model developed in the previous section. In the conservative dynamics, the secular energy given by Hamiltonian (\ref{eq3c}) is a function of the action variable $I^*_1$ (consequently, of $e_1$). The dissipation of the energy occurs when friction forces are introduced and the system loses (or gains) the secular energy during the evolution.

Moreover, the Hamiltonian (\ref{eq3c}) is parameterized by the angular momentum ${\rm AM}$ (through ${\rm AMD}=L_1+L_2-{\rm AM}$) and the ratios $m_2/m_1$ and  $a_1/a_2$. All these parameters may vary under action of some external processes (e. g. planetary migration, mass-loss in close-in planets etc). In particular, the variations of the planetary semi-major axes are generally related to changes of the orbital energy and orbital migration of the planets. However, it is worth noting that, in this section, we do not investigate the dissipation of the orbital energy, defined by the Keplerian motion of the planet. We concentrate our attention on the secular component of the energy given by the Hamiltonian (\ref{eq:Hamilnumeric}), which contains no Keplerian terms (these terms being constants in the secular conservative problem were omitted). Nevertheless the secular Hamiltonian is still dependent on the semi-major axes of the planets (more precisely, on their ratio) as a parameter; thus its change will provoke the changes in the secular dynamics.

To facilitate our understanding, we vary the described above constants separately and study the behaviour of the system in each case.

\subsection{Secular motion with friction}\label{sec2-1}

First, we consider the case of the loss/gain of the secular energy in the system defined by the Hamiltonian (\ref{eq3c}) and moving under action of a friction force. The equations of motion may be written as:
\begin{eqnarray}\label{eq:motion1}
\dot{I}^*_{1} &=& -\frac{\partial {\mathcal E}_1}{\partial w^*} -\epsilon\,f(I^*_{1}) = -\epsilon\,f(I^*_{1})\\
\dot{w}^*&=&\frac{\partial {\mathcal E}_1}{\partial I^*_{1}}=2\beta,
\label{eq:motion}\end{eqnarray}
where we assume a general form of the $w^*$--independent non-Hamiltonian term $f(I^*_{1})$, with a coefficient of friction $\epsilon$. The function $f(I^*_{1})$ is so-called Rayleigh's  dissipation function, given by a positive definite quadratic form in the time derivative of the coordinate, and represent frictional forces which are proportional to velocities (for details see Mierovich 1970, Chapt. 2). The coefficient $\epsilon$ has either positive values, in the case of the loss of energy in the system, or negative values, when the system gains energy. It is worth emphasizing that the friction force acts on the variable $I^*_{1}$ ($e_1$) alone, while the parameters (the angular momentum, its deficit and the mass and semi-major axes ratios) remain constant during the evolution. Since ${\rm AMD}$ and $2\beta$ are constant, the motion of the damped system occurs on the Pauwels' sphere of the unchanged radius.

As shown in Section \ref{sec1-2}, the secular system is linear on a sphere; thus its behaviour is similar to well known behaviour of the harmonic oscillator moving under acting friction force (see for details Andronov et al. 1966, Chapt. I). For $\epsilon=0$, the energy is conserved and  the solution of the system (\ref{eq:motion1})-(\ref{eq:motion}) is trivial: $I^*_{1}={\rm const}$ and the angle $w^*$ is obtained through a simple quadrature as $w^* =2\beta\,t + const$.

\begin{figure}
\def\capfrac{1}
\centerline{\hbox{
\epsfig{figure=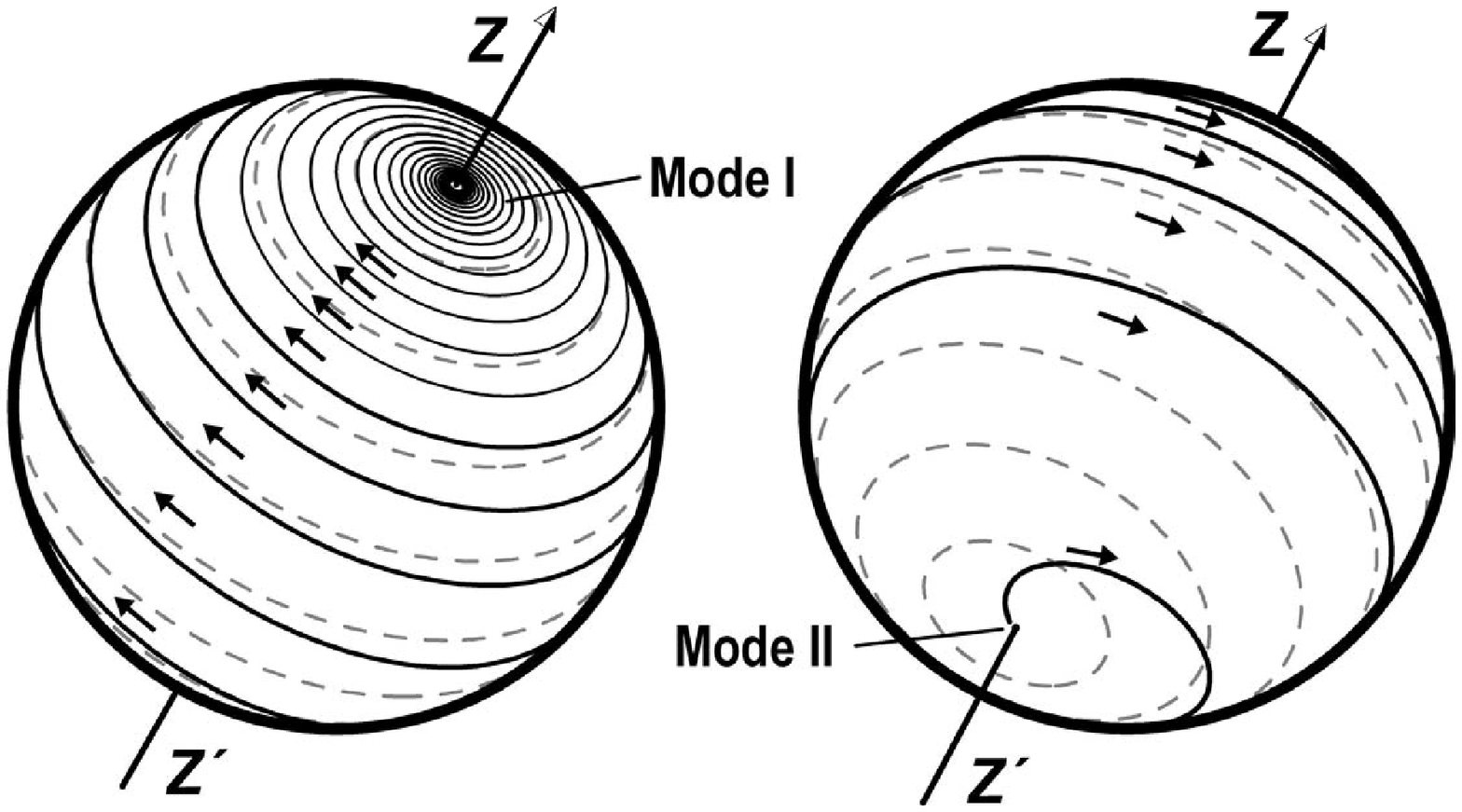, height=5cm,angle=0}
}}
\vspace{0.5cm}
\centerline{\hbox{
\epsfig{figure=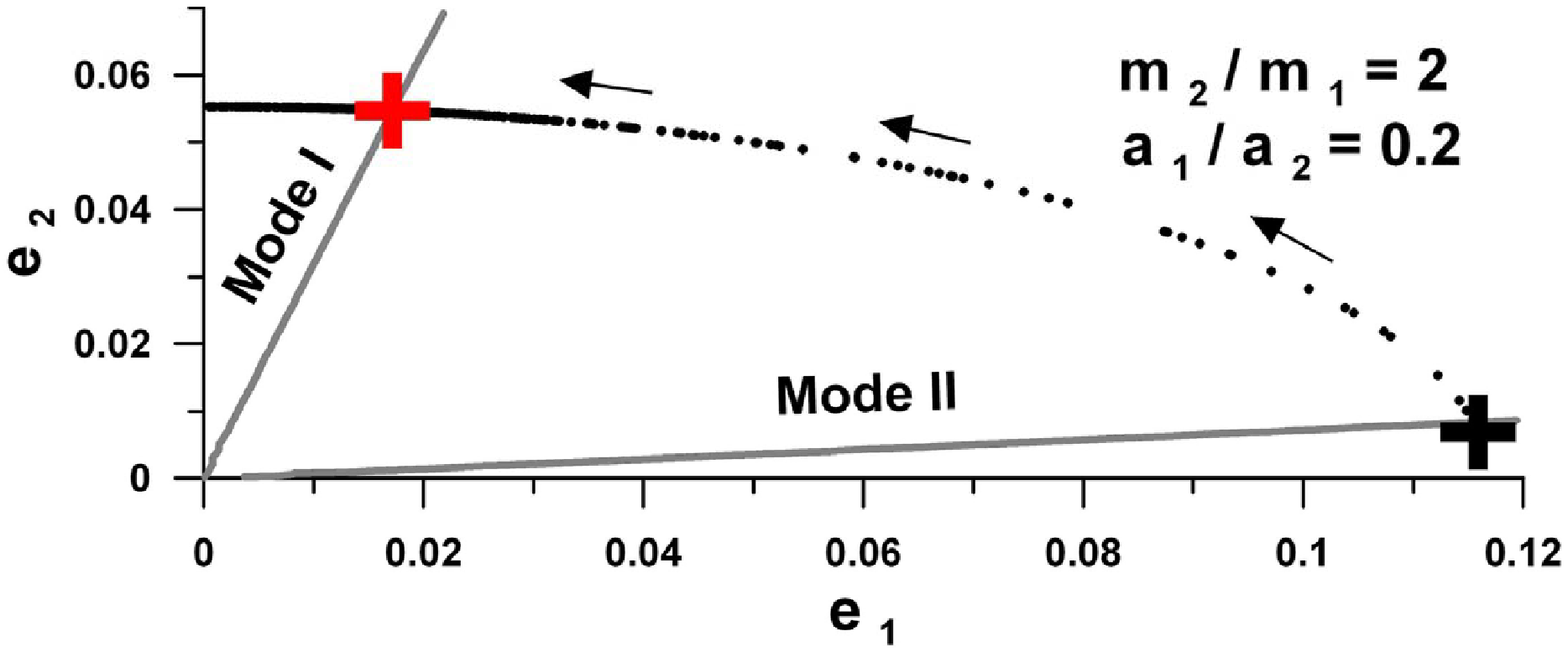, height=3.5cm,angle=0}
}}
\centerline{\hbox{
}}
\vspace{0.5cm}
\centerline{\hbox{
}}
\caption{{\it Top}: Schematic presentation of a damped oscillatory process, which takes place when the dissipation rate is slower than the proper oscillation period. The spiral-like trajectory of the secular system is winding to Mode I of motion (left graph), which plays role of the stable focus. On the right graph, the other side of the Pauwels' sphere shows the same trajectory, unwinding from Mode II, which is an unstable focus.  The dashed curves are circular trajectories of the conservative system shown in Figure \ref{bolas}. {\it Bottom}: the same trajectory shown in the top graphs, on the ($e_1$,$e_2$)--plane. The gray thick lines are the families of conservative stationary solutions, Mode I and Mode II. The system starts at the configuration show by a black symbol and is stopped at configuration shown by a red symbol. Arrows show the direction of the drift of the dissipative motion.
}
\label{figure4}
\end{figure}

If friction is positive ($\epsilon > 0$), $I^*_{1}$ is decreasing to its minimal (zero) value. If dissipation rate is slower than the proper period of the system, the decrease of $I^*_{1}$ provokes damped oscillations of the eccentricity $e_{1}$. All possible trajectories on the Pauwels' sphere, which were simple circles in the conservative case (see Figure \ref{bolas}), have now the form of a spiral; one of these trajectories calculated for the system (\ref{eq:motion1}) is shown in Figure \ref{figure4}\,{\it top}. All together, they form a family of spirals, enclosed in each other, with an asymptotic point at the equilibrium of the system (\ref{eq:motion1}). A singular point of this kind is known as a focus.

The stability of a focus is related to whether the trajectories in its close vicinity are winding (incoming) or unwinding (outgoing) with respect to the direction of motion. If the orbits are incoming, the focus is stable. For example, on the Pauwels' sphere in Figure \ref{figure4}\,{\it left}, the stable focus is close to the Mode I of motion. Otherwise, moving along a spiral orbit, the system moves away from the Mode II (Figure \ref{figure4}\,{\it right}),  in direction of Mode I; in this case, Mode II is an unstable focus. The same trajectory projected on the ($e_1$,$e_2$)--plane is shown in the bottom panel in Figure \ref{figure4}. The evolution of the system on this plane is confined to the level of the constant $\rm{AMD}$: it starts in the vicinity of the Mode II stationary solution (black cross) and is stopped when the system reaches the Mode I stable configuration (red cross).

In the case of the negative friction, the system gains energy. There will be no longer the damping but the reinforcement of the oscillations. The variable $I^*_{1}$ is increasing to its maximal ($\rm{AMD}$) value. Portrait on the phase space  is still a family of spirals, however, the direction of the spirals is inverted with respect to that obtained for systems with a positive friction.

It is worth noting that, in contrast with the dynamics of the harmonic oscillator, the increasing with time deviation of the secular system from the unstable focus (Mode II in Figure \ref{figure4}) has an upper limit, given by the position of Mode I of motion. This is due to the fact that the phase space of secular systems is a sphere; thus all possible trajectories, unwinding the unstable focus, will converge to the stable one. As a consequence, the dissipative system will ultimately evolve into a steady state.

Finally, if friction is positive ($\epsilon > 0$) and its rate is faster then the proper period of the system, we have an aperiodic damping. This case appears in our simulations and will be discussed in Section 5.
\begin{figure}
\def\capfrac{1}
\centerline{\hbox{
\epsfig{figure=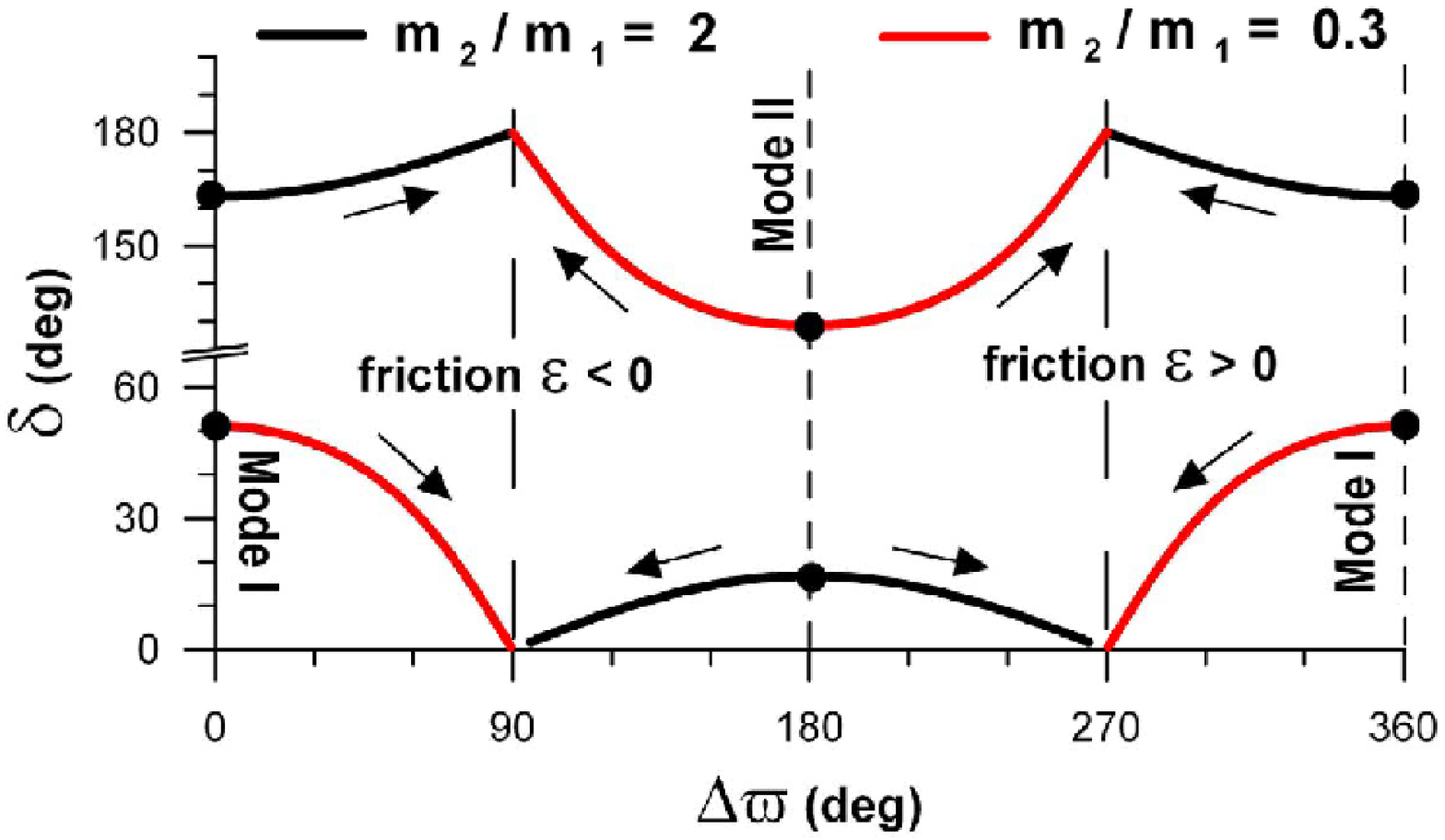, height=5cm,angle=0}
}}
\caption{Stationary solutions of the dissipative system (\ref{eq:motion1})-(\ref{eq:motion}) on the plane ($\Delta\varpi$,$\delta$). Large dots show location of the solutions of the conservative problem when $\epsilon=0$. The left half-plane is characterized by the negative $\epsilon$--values, while the right half-plane by the positive $\epsilon$--values.  Arrows show direction of increasing $\epsilon$ (in abs. value). Black curve was obtained for the mass ratio $m_2/m_1=2.0$, while the red curve for $m_2/m_1=0.3$.}
\label{figure5}
\end{figure}

To illustrate the dependence on the friction coefficient, we calculate the equilibria of the system (\ref{eq:motion1})--(\ref{eq:motion}) as a function of $\epsilon$. For this task, we need a specific form of the friction force: for the sake of simplicity, it was chosen to be proportional to the action variable $I^*_{1}$. The coefficient $\epsilon$ was varied in the range from 0 to 1, in the units of the proper frequency of motion, $2\,\beta$.

The solutions are shown in Figure \ref{figure5}, in the plane ($\delta$, $\Delta\varpi$), where $\Delta\varpi$ is the characteristic angle of the secular model and $\delta$ is a function of the partial angular momentum deficit of the inner planet, $I_1$, and, consequently, of the inner eccentricity. The  parameter $a_1/a_2$ of the secular model was fixed at $0.2$ and two values of the parameter $m_2/m_1$ were used: 2.0 and 0.3. For a given $a_1/a_2$ and small eccentricities, the first value of $m_2/m_1$ satisfies the condition $|C| > |D|$  and the second the condition $|C| < |D|$ (see Equation (\ref{condition1})). The critical value of the mass ratio, obtained for $|C| = |D|$ and $a_1/a_2=0.2$, is $m_2/m_1=0.447$.

The black curve in Figure \ref{figure5} shows the foci obtained for  $m_2/m_1=2.0$. Since $|C| > |D|$ in this case, the conservative equilibrium Mode I is in the lower hemisphere of the Pauwels' sphere, where $\delta > \pi/2$, and the opposite Mode II is in the upper hemisphere, where $\delta < \pi/2$. Their positions are shown by large dots on the black curve, at $\Delta\varpi=0/360^\circ$, for Mode I, and, at $\Delta\varpi=180^\circ$, for Mode II.

The conservative stationary solutions (at $\epsilon=0$) are extrema of the functions $\delta(\epsilon)$ and $\Delta\varpi(\epsilon)$. Therefore, for small non-zero values of $\epsilon$, the equilibria of the dissipative system are close to the equilibria of the conservative one. For this reason, for small $\epsilon$, we continue to refer the foci close to $\Delta\varpi=0$ as Mode I and close to $\Delta\varpi=180^\circ$ as Mode II of motion.

Discontinuities which occur at $\Delta\varpi=90^\circ/270^\circ$ and $\delta=0/180^\circ$ clearly separate two modes of motion. From the definitions in Equation (\ref{eq2}), it is easy to verify that theses discontinuities are related to the singularities of the secular problem (\ref{eq:hamil}), which take place at $I_1=I_2=0$ and $I_1=I_2$ (see Equations (10) and (11) in Michtchenko and Ferraz-Mello 2001).

The foci obtained for  $m_2/m_1=0.3$ are shown in Figure \ref{figure5} by the red curve. In this case, $|C| < |D|$ and the location of the equilibria is opposite to that of the previous case: the Mode I is located now in the upper hemisphere ($\delta < \pi/2$), while Mode II is in the lower hemisphere ($\delta > \pi/2$). In the rest, the evolution of the foci with the increasing value of $\epsilon$ is similar; particulary, the discontinuities also occur at $I_1=I_2=0$ and $I_1=I_2$.

Which of the Mode I and Mode II foci will be stable or unstable depends on the sign of $\epsilon$ and the parameters of the system and is discussed in Section \ref{sec2-3}.

\begin{figure}
\def\capfrac{1}
\centerline{\hbox{
\epsfig{figure=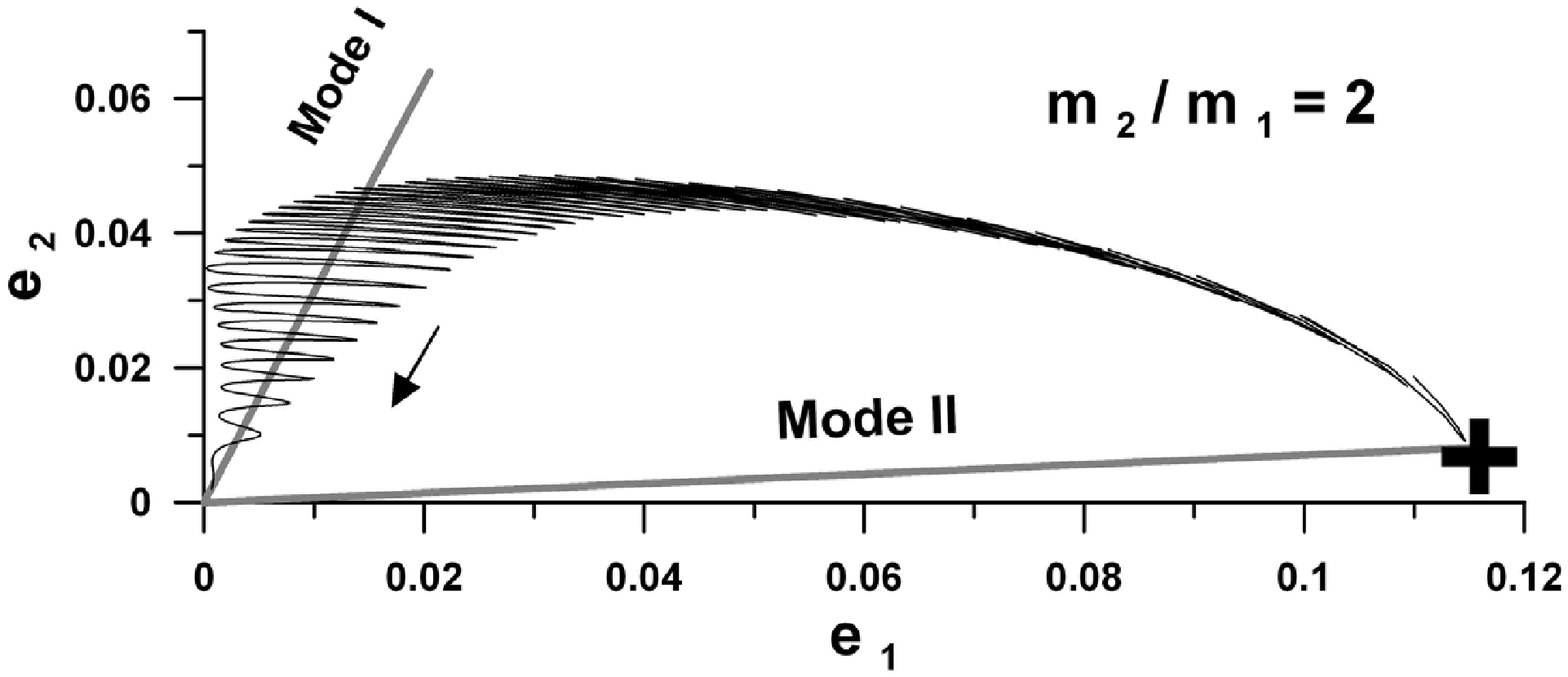, height=4cm,angle=0}
}}
\caption{Trajectory of the secular system with the inner planet in the linear regime, on the ($e_1$,$e_2$)--plane. The gray lines show the families of Mode I and Mode II stationary solutions. The system starts at the configuration show by a black symbol and is stopped when both orbits are circularized. Arrow shows the direction of the drift of the center. The orbit was calculated with the mass ratio $m_2/m_1=2.0$ and the fixed value of $\rm{AM}$.}
\label{figure4-3}
\end{figure}

\subsection{Variation of the semi-major axes ratio and the angular momentum}\label{sec2-2}

Now we consider the dissipative system shown in Figure \ref{figure4}, whose semi-major axes ratio varies, due to some hypothetical processes which keep unchanged the other two parameters,  $m_2/m_1$ and $\rm{AM}$. Although ${\rm AMD}$ is not longer constant, the phase space of the system can be still considered as the Pauwels' sphere, whose radius varies during the evolution, following the changes of ${\rm AMD}$. In such a way, we introduce an 'instantaneous' Pouwels' sphere.

If the $a_1/a_2$--variation is sufficiently slow, the trajectory of the system on the sphere continues to be a spiral approaching a stable center, but the position of the center on the Pauwels' sphere is continuously dislocated together with changes of $a_1/a_2$. Projected on the ($e_1$,$e_2$)--plane, the trajectory of the system is not longer confined to one ${\rm AMD}$--level (as in Figure \ref{figure4}\,{\it bottom}). As shown in Figure \ref{figure4-3}, the damped oscillations of the eccentricities occur now around a center that drifts along the family of Mode I stationary solutions of the conservative case. This family shown by a gray line was calculated using the Hamiltonian (\ref{eq:Hamilnumeric}) of the general model, assuming continuous values of the semi-major axes ratio and the fixed values of the angular momentum and the mass ratio ($m_2/m_1=2.0$).

The trajectory shown in Figure \ref{figure4-3} was obtained through numerical integration of the equations of dissipative motion of the secular system (\ref{eq:motion1})--(\ref{eq:motion}), including the linearly decreasing in time $a_1$. When $a_1$ is decreased (${\rm AMD}$ is also decreased assuming ${\rm AM}=$const), the center of the oscillating system slides down the curve of stationary solutions, in direction of the origin. The evolution is stopped when the system approaches the origin, when both planetary orbits are circularized. At the origin, ${\rm AMD}=0$ and the phase space of the secular system is degenerated in a single point (radius of the Pauwels' sphere tends to zero). As a consequence, we can calculate the minimal possible value of $a_1/a_2$ during the orbital decay (Rodr\'iguez et al. 2011a). Indeed, imposing $e_1=e_2=0$ in Equation (\ref{eq:AM}), we obtain, up to first order in masses,
\begin{equation}\label{eq:a1-min}
\sqrt{ \left(\frac{a_1}{a_2}\right)_{\rm{min}} }=(\rm{AM}^\prime-1)\frac{m_2}{m_1},
\end{equation}
where $\rm{AM}^\prime=\rm{AM}/m_2\sqrt{a_2}$.

When the orbit of the inner planet is expanded, the center slides up the Mode I family, in direction of high eccentricities. In the end, the trajectory will reach a very-high-eccentricity region, where resonant and short-period interactions destabilize the planetary motion.

The behaviour of the dissipative secular system in the case when the angular momentum $\rm{AM}$ varies, but the other parameters are conserved, is essentially same as described in the previous case (see Figure \ref{figure4-3}). The changes of $\rm{AM}$ provokes dissipation of the energy and the angular momentum deficit. As a result, the centers of the secular motion drift on the Pauwels' sphere during the evolution of the system, meanwhile the sphere changes its size. The only difference with the previous case is that the evolutionary track of the system is calculated for all possible values of the angular momentum and the fixed $a_1/a_2$ and $m_2/m_1$.

\subsection{Stability of the centers}\label{sec2-3}

The determination of stability of the centers is immediate when the problem is written in terms of the variable $I^*_1$ defined in Equation (\ref{eq:I1}). Indeed, in this case, $I^*_1$ (and the secular energy) is maximal (minimal) at Mode I (resp. Mode II) of motion (see Section \ref{sec1-2}). Thus, for positive friction when the energy is lost, the system will converge to the Mode II, and, for negative friction, when the system gains energy, to the Mode I.

However, the determination of stability of the centers becomes more complicated when the dissipation is given in terms of eccentricity $e_1$, as in Equation (\ref{condition}). The rigorous way to obtain the solution in this case is to assert the relationship between the variations $\Delta I^*_1$ and $\Delta e_1$, performing inverse  transformation from the variable $I^*_1$ in Equation (\ref{eq:I1}) to the eccentricity $e_1$.

The other way to solve the problem is  based on the definitions done in Section \ref{sec1}.  For instance, from Equations (\ref{condition}) and (\ref{eq2}), we deduce that the orbital decay $\Delta a_1<0$ will produce the following variations:
$$
\Delta e^{\rm{ex}}_1<0 \Rightarrow \Delta I_{1}<0 \Rightarrow \Delta\delta >0.
$$
The above conditions show that, during the orbital decay, the system will evolve into the center with the smaller partial angular momentum  deficit $I_1^Z$, which is located in the lower hemisphere of the instantaneous Pauwels' sphere. On contrary, during the expansion of the orbit of the inner planet ($\Delta a_1>0$), the system will evolve into the center with the larger $I_1^Z$. It is easily verified that the results are equivalent substituting index 1 by the index 2 in the above deduction.

\end{appendix}

\section*{Acknowledgments}
This work has been supported by the Brazilian National Research Council - CNPq, the S\~ao Paulo State Science Foundation - FAPESP. The authors gratefully acknowledge the support of the Computation Center of the University of S\~ao Paulo (LCCA-USP) and of the Astronomy Department of the IAG/USP, for the use of their facilities.

\vspace{-0.5cm}
\section*{\centering { \normalsize REFERENCES}}
\vspace{-0.2cm}

\begin{list}{}{\setlength{\leftmargin}{0.5cm}
		\setlength{\itemindent}{-0.5cm}}
\item
Andronov A.A., Vitt A.A., Khaikin S.E., 1966. Theory of oscillators. Pergamon Press LTD.

\item	
Armitage P.J., 2010, Astrophysics of Planet Formation. Cambridge University Press, Cambridge, UK (further information on this title in: www.cambridge.org/9780521887458).

\item
Batygin K., Laughlin G., Meschiari S., Rivera E., Vogt S., Butler P., 2009, 
ApJ, 699, 23

\item
Beaug\'e, C., Michtchenko T. A., Ferraz-Mello, S., 2006, MNRAS, 
365, 1160 

\item Brouwer, D., and G.M. Clemence 1961. Methods of Celestial Mechanics. Academic Press Inc., New York.

\item	
Callegari N.Jr., Michtchenko T.A., Ferraz-Mello S., 2004, 
Celest. Mech. Dyn. Astr., 89, 201 

\item
Cazenave, A.,  A.  Dobrovolskis, A. and Lago, B., 1980, 
Icarus, 44,  730

\item
Correia A.C.M., Laskar J., 2010, Tidal Evolution of Exoplanets. In: Exoplanets (S. Seager, ed.), Univ. Arizona Press (arXiv:1009.1352)

\item
Fern\'andez J. A., Ip W.-H., 1984, 
Icarus, 58, 109

\item
Fern\'andez J. A., Ip W.-H., 1996, 
Planetary and Space Science, 44, 431

\item
Ferraz-Mello S., Beaug\'e C., Michtchenko T. A., 2003, 
Celest. Mech. Dyn. Astr., 87, 99  

\item
Ferraz-Mello S., Tadeu dos Santos M., Beaug\'e C., Michtchenko T. A., Rodr\'iguez A., 2011, 
A\&A, (in press).

\item
Goldreich P., Sari R., 2003, 
ApJ, 585, 1024

\item
Hadjidemetriou, J. D. \& Voyatzis, G., 2010, 
Celest. Mech. Dyn. Astr., 107, 3

\item	
Kley W., 2000,
MNRAS, 313, L47   

\item	
Kley W., 2003, 
Celest. Mech. Dyn. Astr., 87, 85

\item	
Lee M.H., Peale S.J., 2002,  
ApJ, 567, 596   

\item
Lubow S.H., Ida S., 2010, Planet Migration. In: Exoplanets (S. Seager, ed.), Univ. Arizona Press (arXiv:1004.4137)

\item
Malhotra, R., 1994, 
Celest. Mech. Dyn. Astr., 60, 373

\item
Malhotra, R., 1995, 
AJ, 110, 420

\item
Mardling, R. A., 2007, 
MNRAS, 382, 1768

\item
Meirovitch L., 1970. Methods of analytical dynamics. McGraw-Hill Book Company, New York.

\item	
Michtchenko T.A., Ferraz-Mello S., 2001, 
Icarus, 149, 357  

\item
Michtchenko T.A., Lazzaro D., Ferraz-Mello S., Roig F., 2002, 
Icarus, 158, 343  

\item	
Michtchenko T.A., Malhotra R., 2004, 
Icarus, 168, 237  

\item	
Michtchenko T.A., Beaug\'e C., Ferraz-Mello S., 2008a, 
MNRAS, 387, 747

\item	
Michtchenko T.A., Beaug\'e C., Ferraz-Mello S., 2008b, 
MNRAS, 391, 215

\item
Milani A., Nobili A.M., 1984, 
Nature, 310, 753

\item
Murray N., Paskowitz M., Holman M., 2002, 
ApJ, 565, 608

\item
Nelson, R. P., Papaloizou, J. C. B., 2002, 
MNRAS, 333, L26

\item
Pauwels, T. 1983, 
Celest. Mech., 30, 229

\item
Rodr\'iguez A., Ferraz-Mello, S., Michtchenko T.A., Beaug\'e C., Miloni O., 2011a, 
MNRAS, (submitted).

\item
Rodr\'iguez A., Michtchenko T.A., Miloni O., 2011b, 
Celest. Mech. Dyn. Astr., (submitted).

\item
Tittemore, W., Wisdom, J., 1988, '
Icarus 74, 172

\item
Wu Y., Goldreich P., 2002, 
ApJ, 564, 1024

\end{list}

\end{document}